\documentclass{aa}
\usepackage{graphicx}\graphicspath{{fig/}}
\usepackage{natbib} \bibpunct[, ]{(}{)}{;}{a}{}{,}
\usepackage{amsmath,amssymb}

\newcommand{\rmd}{{\rm d}}
\newcommand{\rme}{{\rm e}}
\newcommand{\bx}{{\mathbf{x}}}
\newcommand{\CA}{{\cal A}}
\newcommand{\CO}{{\cal O}}

\newcommand{\hMpc}{{\ifmmode{h^{-1}{\rm Mpc}}\else{$h^{-1}$Mpc}\fi}}
\newcommand{\intphi}{\int\limits_0^{2\pi} \frac{\rmd\phi}{M_0(B_r)}}
\newcommand{\intx}{\int\limits_{-s}^{1}\rmd x}
\newcommand{\iintphi}{ \int\!\!\!\!\int\limits_{\!\!\!\!0}^{2\pi}\!\!
\frac{\rmd\phi_1\rmd\phi_2}{M_0(B_r)^2} } 
\newcommand{\iintx}{\int\!\!\!\!\int\limits_{\!\!\!\!\!\!\!-s}^{1}\!\! 
\rmd x_1\rmd x_2 }

\begin{document}


\title{Morphological fluctuations of large--scale structure:\\
the PSCz survey}
\author{
M.~Kerscher\inst{1,2},
K.~Mecke\inst{3,4} \and 
J.~Schmalzing\inst{5,2},
C.~Beisbart\inst{2},
T.~Buchert\inst{6,7,2} \&
H.~Wagner\inst{2}}

\offprints{M.~Kerscher,
\email{kerscher@theorie.physik.uni-muenchen.de}}

\institute{
Department of Physics and Astronomy, 
The Johns Hopkins University, Baltimore, MD 21218, USA
\and
Sektion Physik, Ludwig--Maximilians--Universit{\"a}t, 
Theresienstra{\ss}e 37, D--80333 M{\"u}nchen, Germany
\and
Max-Planck-Institut f\"ur Metallforschung, Heisenbergstr. 1, D-70569
Stuttgart, Germany   
\and
Institut f\"ur Theoretische und Angewandte Physik, Fakult\"at f\"ur
Physik, Universit\"at Stuttgart, Pfaffenwaldring 57, D-70569 Stuttgart, Germany
\and
Teoretisk Astrofysik Center, Juliane Maries Vej 30,
DK--2100 K{\o}benhavn \O, Denmark
\and 
Theoretical Astrophysics Division, National Astronomical Observatory
 2--21--1 Osawa Mitaka Tokyo 181--8588, Japan
\and 
D\'epartement de Physique Th\'eorique, Universit\'e de
Gen\`eve, 24 quai E.~Ansermet, CH--1211 Gen\`eve, Switzerland
}

\date{Received: 15 January 2001 / Accepted 24 April 2001}

\abstract{ In a follow--up study to a previous analysis of the IRAS
1.2Jy catalogue, we quantify the morphological fluctuations in the
PSCz survey.  We use a variety of measures, among them the family of
scalar Minkowski functionals.  We confirm the existence of significant
fluctuations that are discernible in volume--limited samples out to
200\hMpc.  In contrast to earlier findings, comparisons with
cosmological $N$--body simulations reveal that the observed
fluctuations roughly agree with the cosmic variance found in
corresponding mock samples.  While two--point measures, e.g.  the
variance of count--in--cells, fluctuate only mildly, the fluctuations
in the morphology on large scales indicate the presence of coherent
structures that are at least as large as the sample.
\keywords{large--scale structure of Universe -- Cosmology: observation
-- Galaxies: statistics}
}

\titlerunning{Morphological fluctuations: the PSCz survey}
\authorrunning{M.~Kerscher et al.}

\maketitle


\section{Introduction}

Nowadays, one of the primary goals of cosmology is the verification of
the plethora of models that have emerged over the years.  To that end,
experiments are being conducted and planned that will ultimately yield
deep and wide galaxy surveys and high--resolution maps of the Cosmic
Microwave Background.  In parallel to the preparation of these large
datasets, statistical methods that are both easy to interpret and
efficient to implement are being developed.

A few years ago, the Minkowski functionals (MFs) were introduced into
cosmology by {\cite{mecke:robust}}.  The MFs can be calculated from
contemporary datasets rather efficiently, and consequently numerous
applications (see {\citealt{kerscher:statistical}} and references
therein, vector--valued MFs are described in
{}\citealt{beisbart:morphometry}) have established the MFs as a
suitable tool for quantifying the morphological properties of the
large--scale structure in the Universe . Apart from their practical
advantages, MFs provide a mathematically well--founded systematic
framework for the study of statistical morphology and contain other
higher--order statistics that were employed previously.

The present work extends the scope of an earlier paper on the
morphological fluctuations in the IRAS 1.2Jy catalogue of infrared
galaxies {\citep{kerscher:fluctuations}}, by using its much deeper
successor, the recently compiled PSCz survey {\citep{saunders:pscz}}.
However, we do not simply repeat the previous study with the enlarged
dataset, but devise additional sensible tests in order to assess the
quality of the data and to compare the observational data to $N$--body
simulations.  {\cite{kerscher:fluctuations}} detected, quite
unexpectedly at that time, significant fluctuations in the
morphological properties between the northern and southern hemispheres
of the IRAS 1.2Jy catalogue in volume--limited samples of up to
200\hMpc\ depth.  While these findings were shown to be significant
and not due to deficiencies of the data, simulations could not
reproduce them, even though they were specifically designed for that
purpose.  Meanwhile, simulations have improved, mainly with respect to
the size of the simulation box.  We shall demonstrate that the large
fluctuations seen in the IRAS 1.2Jy catalogue can still be found in
the PSCz survey.  However, state--of--the--art simulations of Cold
Dark Matter (CDM) scenarios are now capable of predicting sufficient
cosmic variance.  Another analysis of the PSCz sample with Minkowski
functionals by {\citet{basilakos:pscz}} focused on the shape of
super--clusters.

This article is organised as follows: Sect.~\ref{sec:morphometry}
briefly explains how Minkowski functionals are calculated from the
PSCz dataset and summarises our findings in a qualitative way.  It
also contains a discussion of possible systematic effects and results
obtained with other statistical methods.  Sect.~\ref{sec:comparison}
compares the observational data with an analytical model and a set of
mock catalogues.  We summarise and provide an outlook in
Sect.~\ref{sec:summary}.  Finally, a few necessary mathematical
results are given in the Appendix.

\section{Measuring the morphology of the PSCz survey}
\label{sec:morphometry}

\subsection{Minkowski functionals}

\begin{table}
\begin{center}
\caption{
\label{table:mingeom}
Minkowski functionals in three--dimensional space expressed in terms
of more familiar geometric quantities.}
\begin{tabular}{cl|ccc}
& geometric quantity & $\mu$  & $M_\mu$ & $M_\mu(B_r)$ \\[1ex]
\hline
$V$ & volume & 0 & $V$  & $4\pi r^3/3$ \\ 
$A$ & surface area &  1 & $A/8$  & $\pi r^2/2$ \\
$H$ & mean curvature & 2 & $H/(2\pi^2)$ & $2 r/\pi$ \\ 
$\chi$ & Euler characteristic & 3 &  $3\chi/(4\pi)$ & $3/(4\pi)$ \\
\end{tabular}
\end{center}
\end{table}

In order to quantify the morphology within the PSCz survey
{\citep{saunders:pscz}} with Minkowski functionals, we interpret the
redshift space positions $\{\bx_i\}_{i=1}^N$ of the $N$ galaxies in
the sample as a realisation of a stationary point process.  Adopting
the Boolean grain method of {\citet{mecke:robust}}, we decorate each
point $\bx_i$ with a sphere $B_r(\bx_i)$ of radius~$r$ and consider the
union set $\CA_r=\bigcup_{i=0}^NB_r(\bx_i)$.
{\citet{hadwiger:vorlesung}} has shown that in three--dimensional
space the four Minkowski functionals $M_\mu(\CA_r)$, with
$\mu=0,1,2,3$ provide a complete morphological characterisation of the
body $\CA_r$.  Table~\ref{table:mingeom} summarises the
interpretations of these functionals in terms of well--known
geometrical and topological quantities.

Reduced, dimensionless Minkowski functionals $\Phi_\mu(\CA_r)$ can be
constructed by normalising with the Minkowski functionals $M_\mu(B_r)$
of a single sphere,
\begin{equation}
\label{eq:Phi-def}
\Phi_\mu(\CA_r) = \frac{M_\mu(\CA_r)}{NM_\mu(B_r)}.
\end{equation}
The mean number density $\rho$ can be estimated in a sample by
$N/|\Omega|$, where $|\Omega|$ is the volume of the sample.
For a Poisson process these functionals can be calculated analytically
{\citep{mecke:euler}} as functions of $\eta=M_0(B_r)\rho$, the
expected number of galaxies in a sphere of radius $r$:
\begin{equation}\label{eq:Poisson}
\begin{array}{rl}
\Phi_0^{\rm P} & = \left(1 - \rme^{-\eta}\right)\ \eta^{-1}, \\
\Phi_1^{\rm P} & = \rme^{-\eta} ,\\
\Phi_2^{\rm P} & = \rme^{-\eta}\ (1 - \frac{3 \pi^2}{32} \eta ), \\
\Phi_3^{\rm P} & = \rme^{-\eta}\ (1 - 3 \eta + \frac{3 \pi^2}{32} \eta^2 ). 
\end{array}
\end{equation}
Obviously, the Minkowski functionals $\Phi_\mu^{\rm P}(\CA_r)$,
$\mu=1,2,3$, contain a damping factor $\rme^{-\eta}$.  A similar
exponential decay is also found for more general cluster processes.
To facilitate the comparison between the data and different models on
large scales, we reduce the functionals by this factor and consider
the quantities
\begin{equation}\label{eq:phi-def}
\phi_\mu(\CA_r)\equiv\frac{\Phi_\mu\left(\CA_r\right)}{\Phi_1^{\rm P}(r)}
\end{equation}
throughout  the following analysis.
\\
Effective methods to calculate the Minkowski functionals for empirical
data points are described in {\cite{mecke:robust}}. To correct for
boundary effects without applying unnecessary statistical assumptions
we proceed as \citet{schmalzing:minkowski} (see also
\citealt{mecke:euler,kerscher:abell}).

\subsection{Selection of the samples}

A detailed description of the PSCz galaxy catalogue may be found in
{\citet{saunders:pscz}}.  For our analysis we extract volume--limited
samples (Table~\ref{table:samples}) from the PSCz survey within the
standard masked area.  The velocities of the galaxies in the local
group frame ($v=v_{\rm hel}+\sin(l)\cos(b)\ 300{\rm km/s}$) are
converted into luminosity distances assuming $q_0=0.5$ and
$\Lambda=0$.
\\
\begin{table}
\caption{
\label{table:samples}
Volume--limited samples from the PSCz survey. $N$ denotes the number
of galaxies.}
\begin{center}
\begin{tabular}{c|c|c|c}
depth [\hMpc] & flux limit [Jy] & $N_{\rm north}$ & $N_{\rm south}$ \\[1ex]
\hline
100 & 0.6 & 1119 & 1078 \\
100 & 0.8 &  676 &  661 \\
100 & 1.2 &  311 &  332 \\
200 & 0.8 &  337 &  323 \\
\end{tabular}
\end{center}
\end{table}
In  order to simplify  the boundary  corrections, we  approximate the
sample  geometry  by  a  spherical  cap  with  galactic  latitude
$b\ge5^\circ$ for the northern  sample and with $b\le-5^\circ$ for the
southern  part. Since by doing so,  we neglect  some  regions which  were
excluded due to galactic absorption or confusion in the IRAS PSC maps,
we filled these empty regions  with random points with the same number
density as in the fully  sampled region.  Differences in the Minkowski
functionals  between  the  filled  and  unfilled  samples  are  barely
visible.

Another issue to be discussed is the flux limit of the PSCz
survey. Tentatively, we evaluated the Minkowski functionals for a
series of volume--limited samples with 100\hMpc\ depth , but with
varying limiting flux.  Only for limiting fluxes above 0.8Jy the
Minkowski functionals stabilise.  Likewise, {\citet{tadros:spherical}}
find stable results only for a flux limit larger than 0.75Jy.
Therefore, we shall adopt a flux limit 0.8Jy for all of our analyses.

This kind of flux dependence does not show up in two--point measures.
The two--point correlation function $\xi_2$, e.g., does not change
significantly when the flux cut is varied. Even the mark correlation
functions {\citep{beisbart:luminosity,szapudi:correlationspscz}} do
not show any luminosity--dependent clustering at the two--point level.
Further   tests   on   selection   effects   will   be   discussed   in
Sect.~\ref{sec:selection-effects}.

\subsection{Fluctuating morphology}

\begin{figure*}
\begin{center}
\includegraphics[width=5.9cm]{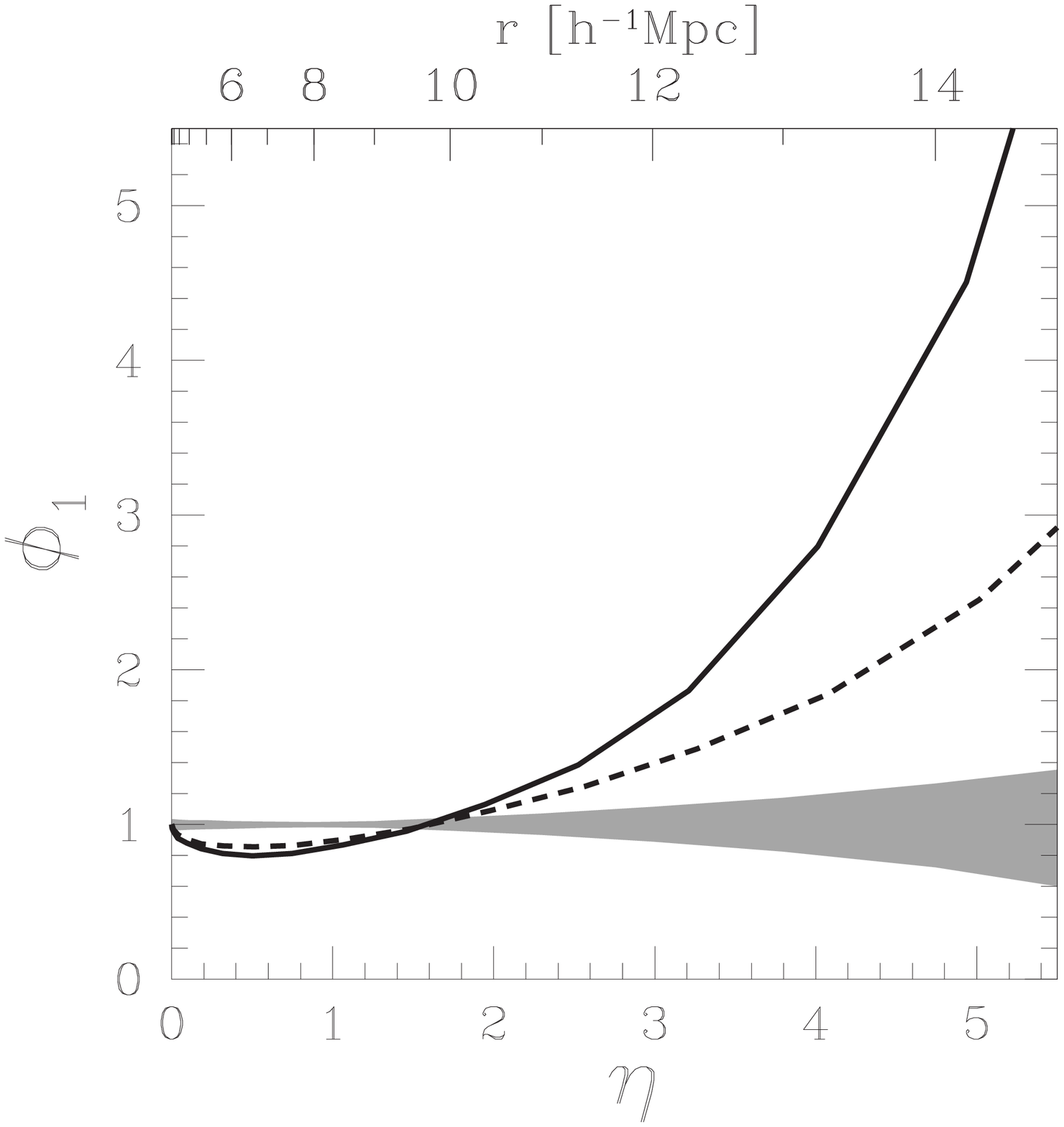}
\includegraphics[width=5.9cm]{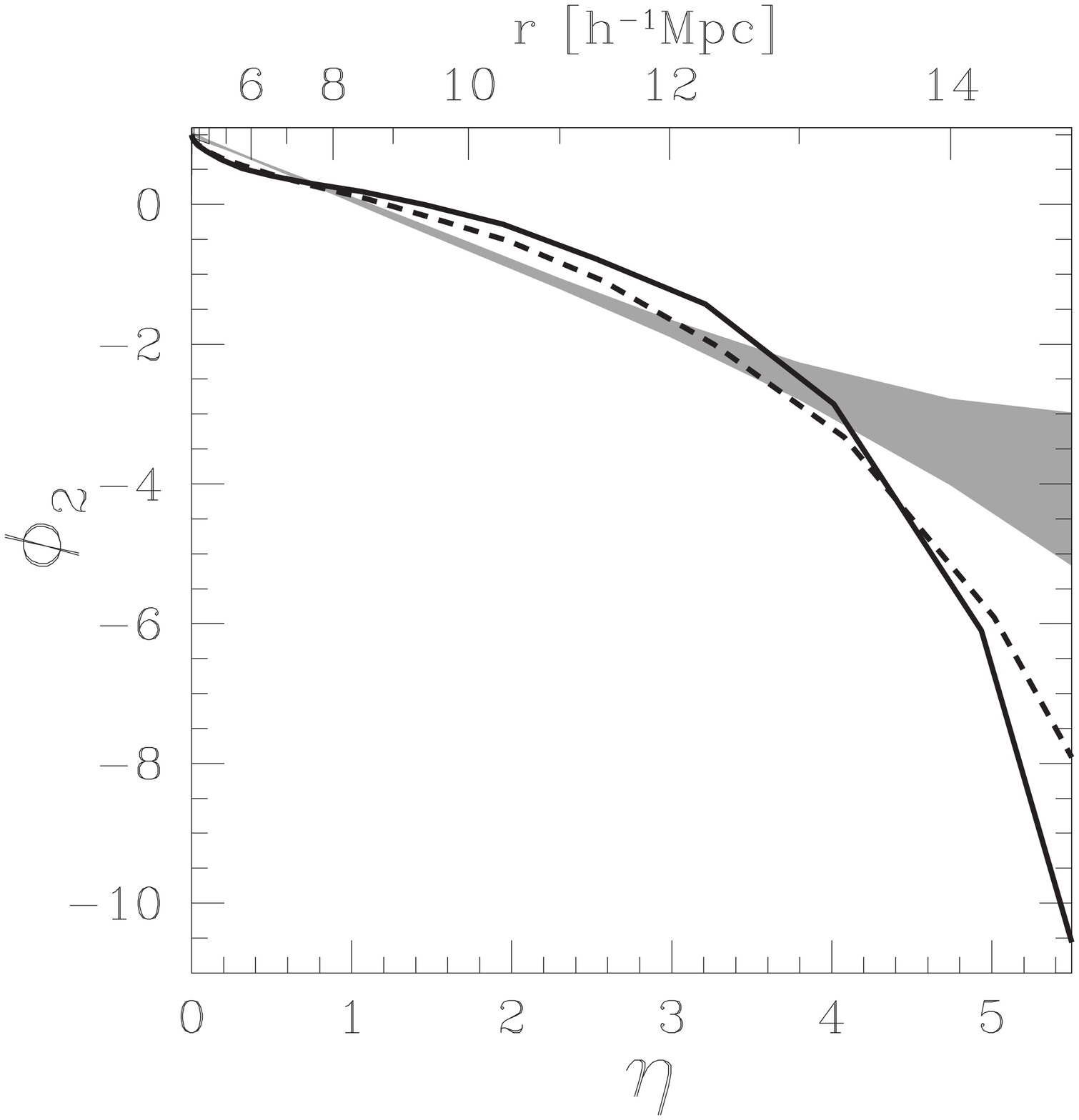}
\includegraphics[width=5.9cm]{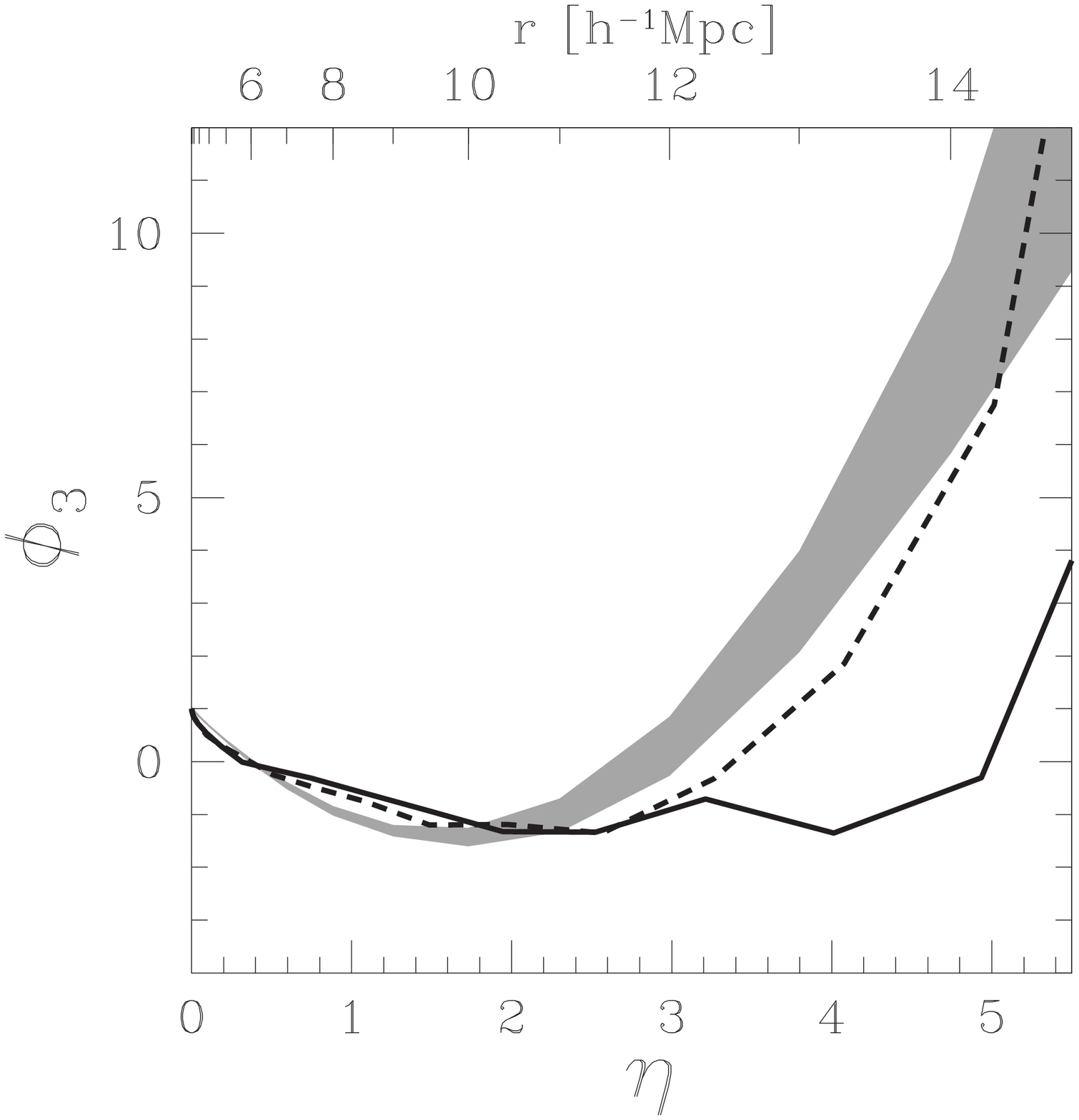}
\end{center}
\caption{
\label{fig:MF-v100}
The MFs of volume--limited samples from the PSCz survey with 100\hMpc\
depth. The southern (solid) and northern parts (dashed) are plotted
separately, and for comparison the one--sigma fluctuations (shaded
regions) for a Poisson process with the same number density are also
shown.  }
\end{figure*}

Fig.~\ref{fig:MF-v100} displays the values of the reduced Minkowski
functionals $\phi_\mu$ as functions of the $\eta$ for the southern and
northern parts of the PSCz survey at a limiting depth of $100\hMpc$.
For comparison, we also show the expectation values for a Poisson
process with the same number density.

In both parts of the sample the galaxy clustering is noticeably
stronger than in the case of randomly distributed points.  Moreover,
the northern and southern parts differ significantly in their
morphological features, the northern part being less lumpy.  The
most conspicuous features are the enhanced surface area $\phi_1$ in
the southern part on scales larger than 10\hMpc, and the decrease of
the Euler characteristic $\phi_3$, which sets in at 12\hMpc.  On scales
above 10\hMpc, the integrated mean curvature $\phi_2$ and the Euler
characteristic $\phi_3$ are negative, indicating an interconnected system
of tunnels.  In the southern part $\phi_3$ remains negative out to
large scales, and the large--scale structure is dominated by
interconnected tunnels contrary to the northern part, where completely
enclosed voids yield positive contributions to the Euler
characteristic.

The morphological fluctuations observed in the PSCz survey appear
comparable to those found in the IRAS 1.2Jy catalogue
{\citep{fisher:irasdata,kerscher:fluctuations}}.  To verify this, we
extract volume--limited samples from the PSCz survey with a limiting
flux of 1.2Jy.  For this comparison we use the Euclidean redshift
space distance as in our original analysis of the IRAS 1.2Jy
catalogue.  In the southern part, we find no difference in the
Minkowski functionals.  The northern part of this PSCz sample,
however, contains less galaxies than the corresponding IRAS 1.2Jy
subsample, probably due to the more restrictive selection criteria
imposed on the PSCz survey.  But still, the MFs of the randomly
sub--sampled northern part of the IRAS 1.2Jy nearly overlap with the
PSCz sample.  Hence we are confident that the MFs pick up the
underlying large--scale structure, which should of course be the same
in both catalogues.

\subsection{Selection effects and deeper samples}
\label{sec:selection-effects}

To strengthen our claims regarding the morphological fluctuations, we
have to exclude some possible sources of error.  To begin with, an
analysis of the 100\hMpc\ sample with a more conservative cut at
$|b|\ge10^\circ$ instead of $5^\circ$ leads to nearly identical
results for the Minkowski functionals.

To check whether our results are influenced by some peculiar alignment
of points, we considered a subsample with only 90\% of the galaxies
randomly chosen from the volume--limited sample. Repeating this
jackknife procedure, we calculate the mean and the error of the
Minkowski functionals.  Still the same prominent fluctuations between
north and south can be seen.  These errors are smaller than the
fluctuations in the random point set.  We want to emphasise that this
jackknife error can only serve as an internal consistency check; it
does not give a reliable estimate of the fluctuations in the
underlying galaxy distribution {\citep{snethlage:bootstrap}}.

Moreover, we select subsamples with ``warm'' and ``cold'' galaxies as
determined from the flux ratio $\frac{f_{100}}{f_{60}}\le2$ and
$\frac{f_{100}}{f_{60}}>2$, respectively, with $f_{100}$ the flux at
100$\mu$m and $f_{60}$ the flux at 60$\mu$m.  In both samples we find
fluctuations between north and south, comparable to those in the
combined sample. Finally, the strength of the fluctuations hardly
diminishes if we consider the measured velocities without a
heliocentric correction.

\begin{figure*}
\begin{center}
\includegraphics[width=5.9cm]{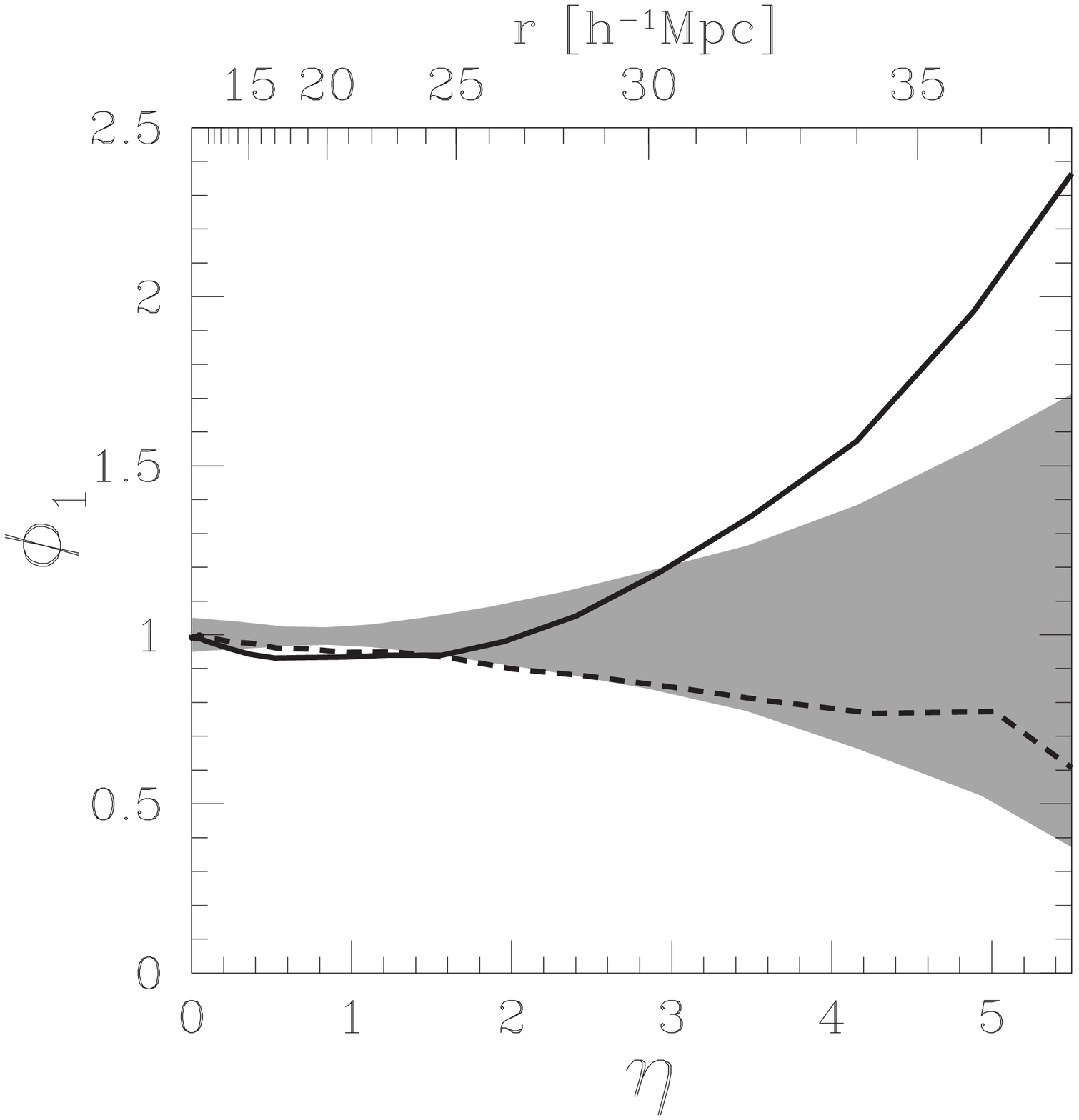}
\includegraphics[width=5.9cm]{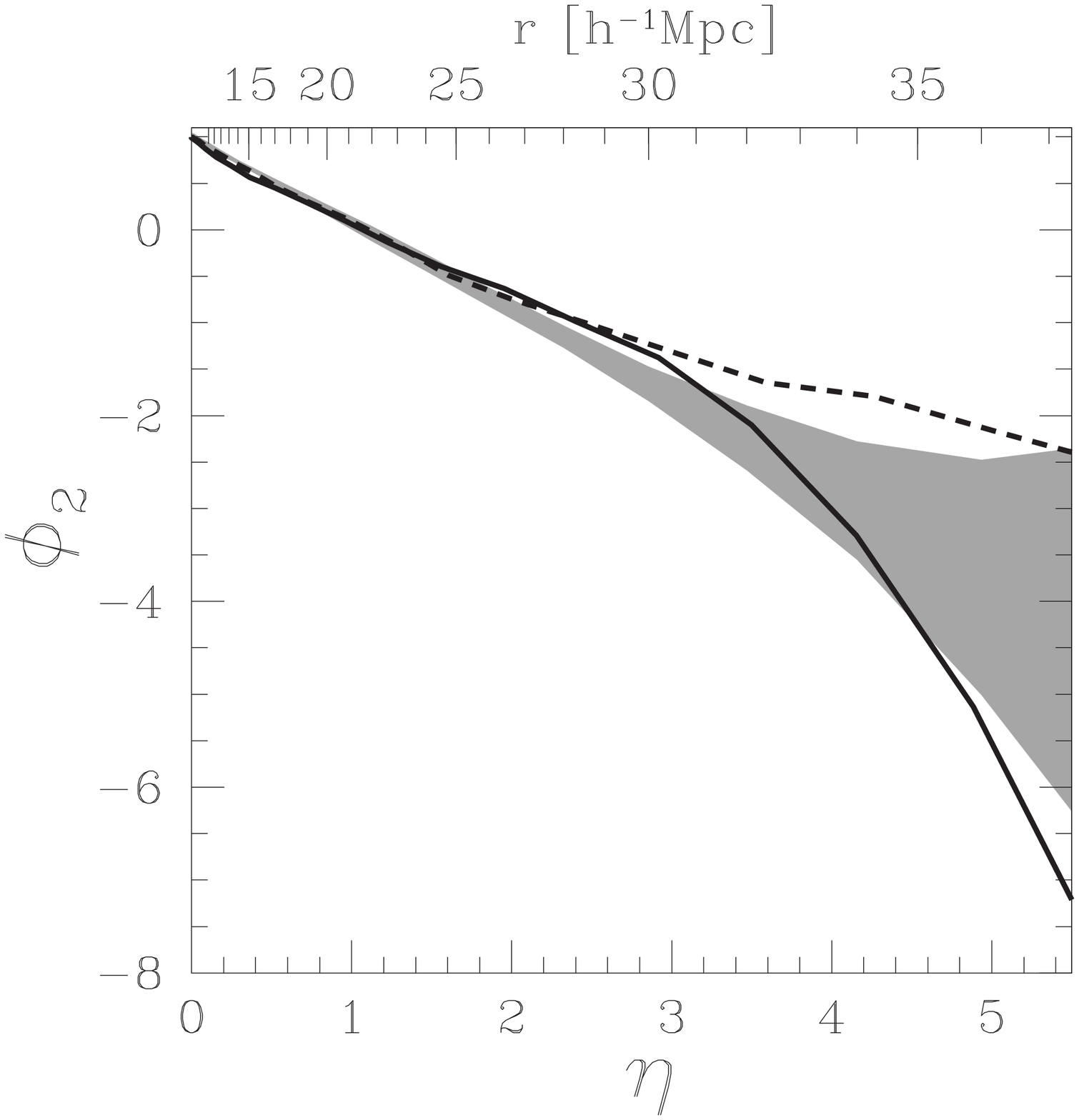}
\includegraphics[width=5.9cm]{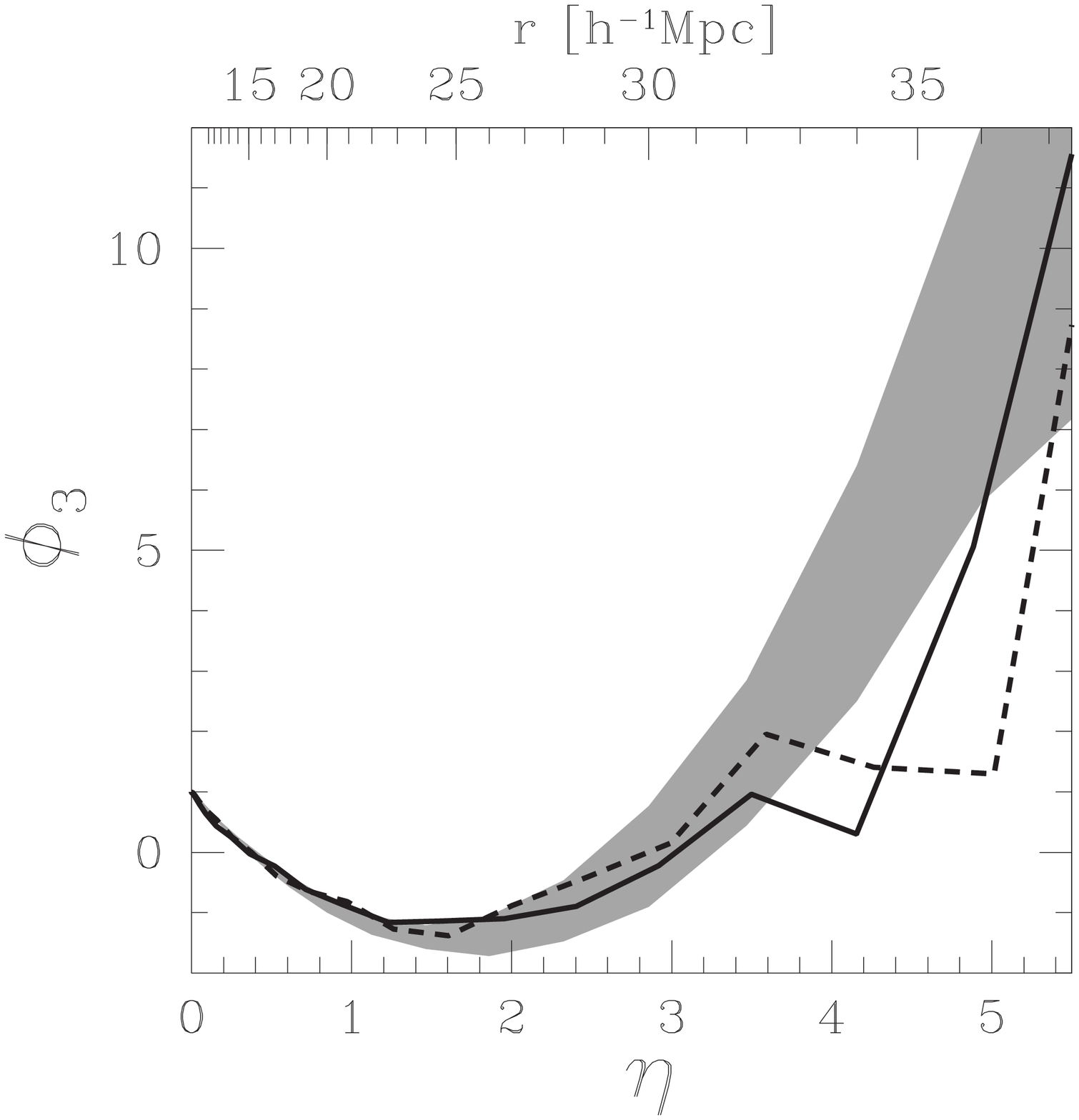}
\end{center}
\caption{
\label{fig:MF-v200}
The Minkowski functionals of volume--limited samples with 200\hMpc\
depth.  See Fig.~\ref{fig:MF-v100} for an explanation of plotting
styles.}
\end{figure*}

To see whether these fluctuations persist on even larger scales, we
investigate volume--limited samples with 200\hMpc\ depth
(Fig.~\ref{fig:MF-v200}).  On small scales, both in the northern and
the southern part, the MFs indicate clustering.  Again and in
particular on large scales, the MFs of the southern part differ from
the MFs of the northern part, showing approximately the same pattern
as in the volume--limited sample with 100\hMpc.  The fluctuations are
now only slightly larger than the one--sigma fluctuations observed in
a Poisson process with the same number density.  As already
demonstrated by {\citet{kerscher:fluctuations}}, the morphology of
such a sparsely sampled catalogue tends to resemble a pure random
sample as dilution increases (see also Sect.~\ref{sec:small-large}).
The  Minkowski   functionals  in  the   volume--limited  samples  with
150\hMpc\  depth give  comparable results,  showing fluctuations  of a
strength between the fluctuations of the 100\hMpc\ and the 200\hMpc\ sample.

\begin{figure*}
\begin{center}
\includegraphics[width=5.9cm]{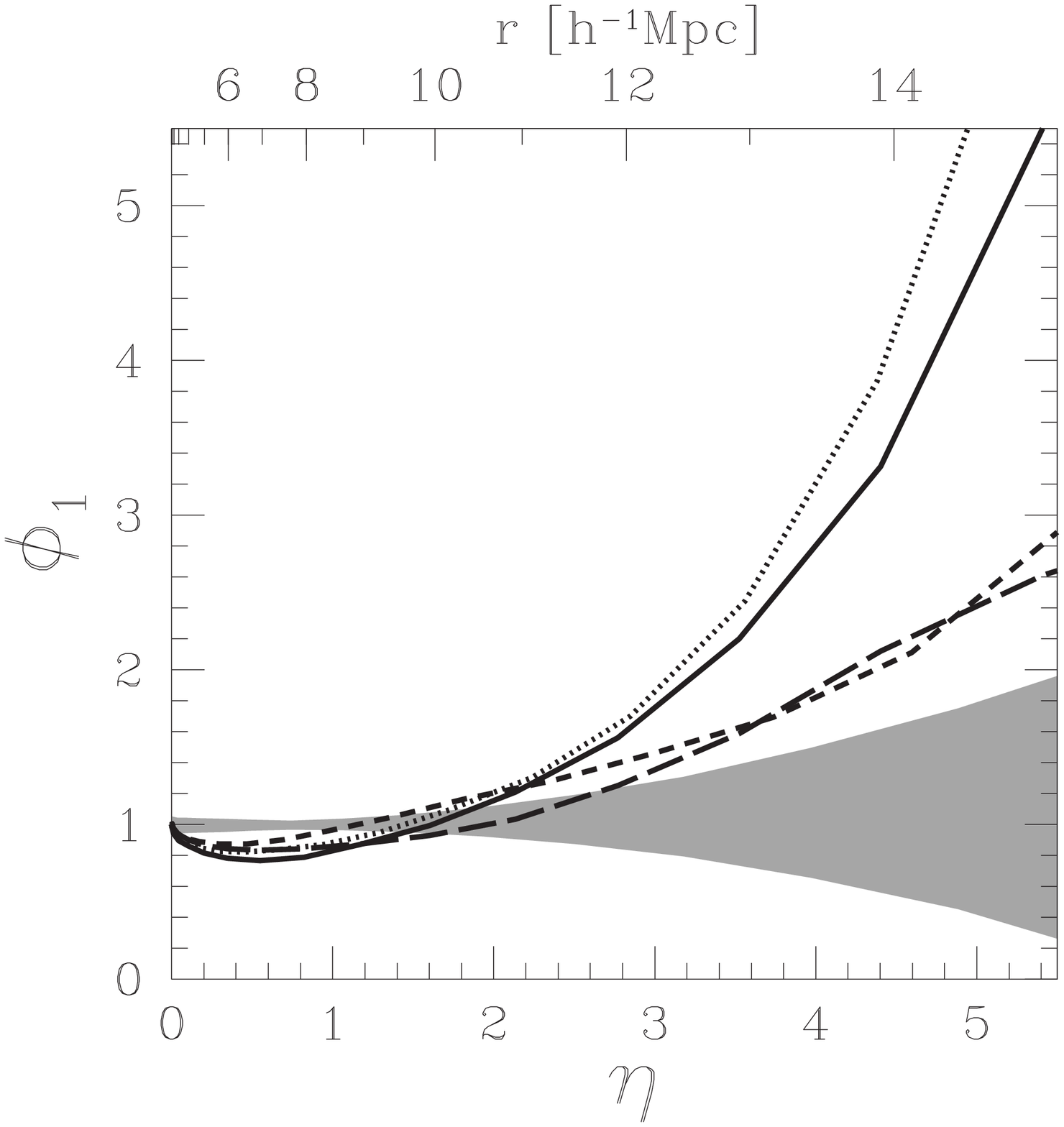}
\includegraphics[width=5.9cm]{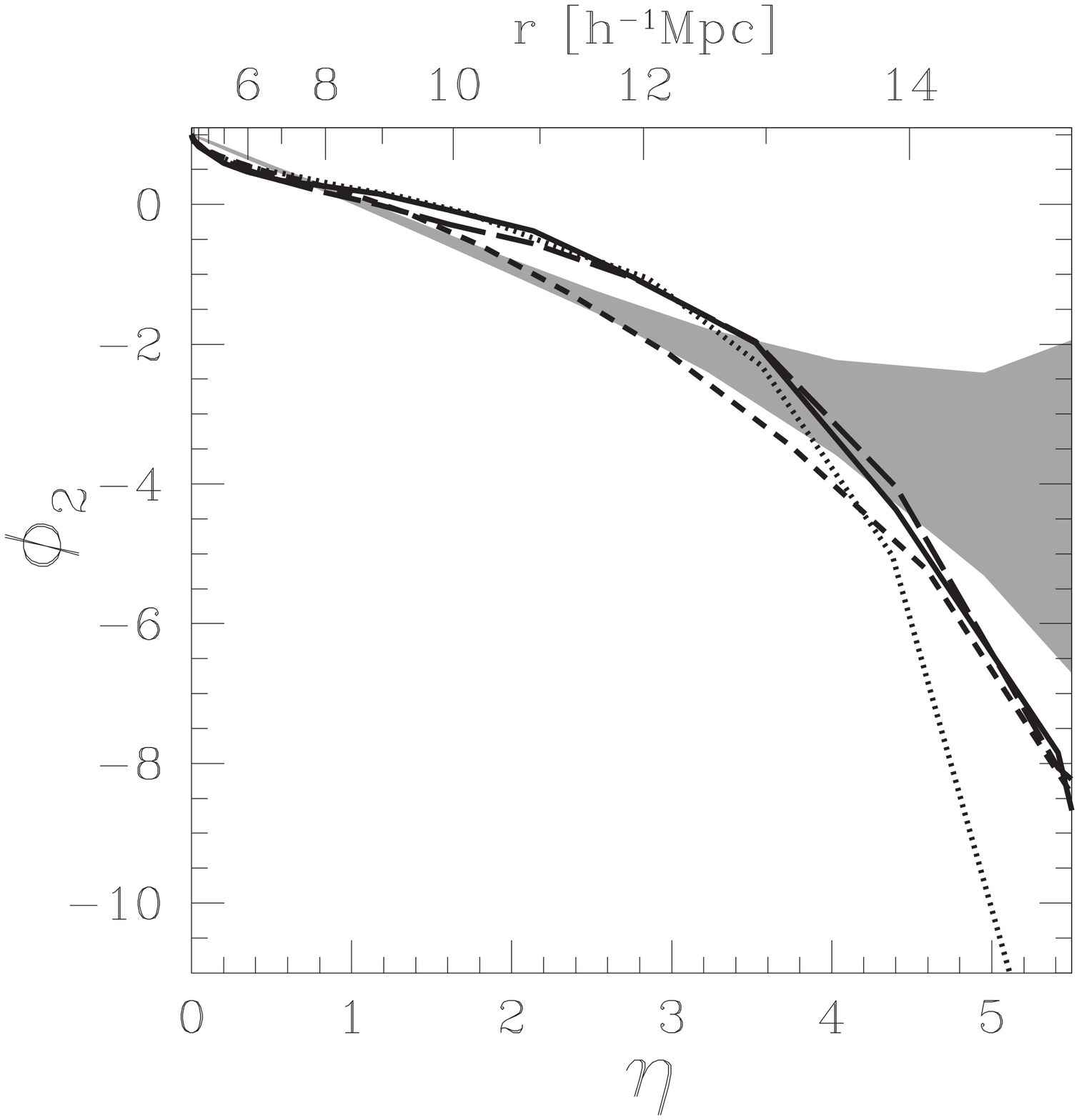}
\includegraphics[width=5.9cm]{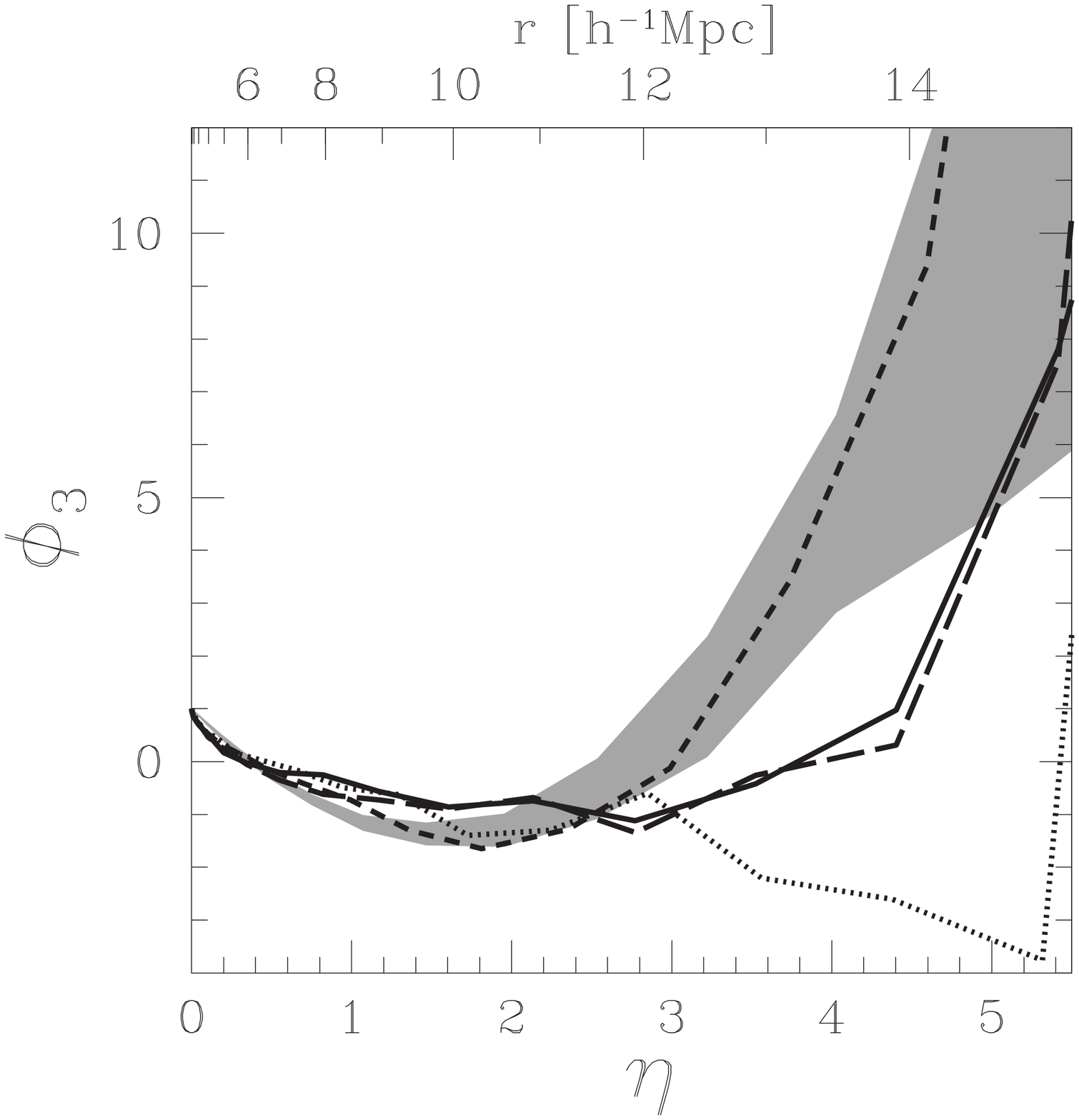}
\end{center}
\caption{
\label{fig:MF-v100-quart}
The MFs of volume--limited samples with 100\hMpc\ depth from the PSCz
survey.  Four different parts of the sample are considered separately:
southern left (solid), southern right (dotted), northern left (short
dash), northern right (long dashed).  As before, the area shows a
Poisson process.  }
\end{figure*}

In {\citet{kerscher:significance}} we could show that the
morphological discrepancies in the 1.2Jy galaxy catalogue are not a
special north--south anisotropy. We proceed similar by cutting both
the northern and the southern part along the $yz$--plane (in galactic
coordinates) into two pieces with the same volume. The Minkowski
functionals of the four samples are shown in
Fig.~\ref{fig:MF-v100-quart} confirming that the fluctuations are
not a peculiar north--south anisotropy, but actually reflect the
generic cosmic variance.

\subsection{Other measures}

\begin{figure}
\begin{center}
\includegraphics[width=7cm]{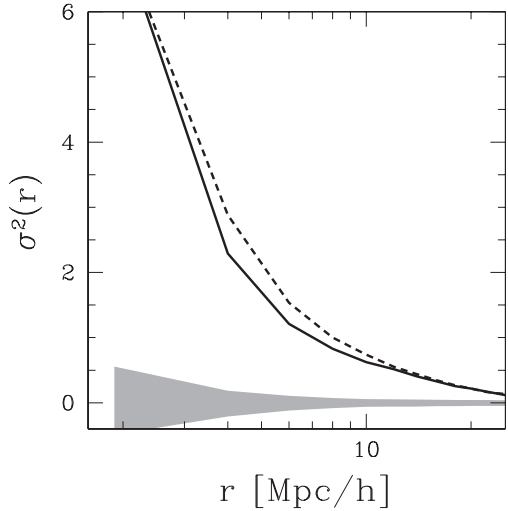}
\end{center}
\caption{
\label{fig:sigma-v100}
Fluctuation of count--in--cells $\sigma^2(r)$ of the volume--limited
sample with 100\hMpc\ depth from the PSCz survey.  Plotting styles are
the same as in Fig.~\ref{fig:MF-v100}. }
\end{figure}

Both from a physical and a methodological point of view, it is
interesting to compare the performance of the MFs with other
statistics.  For instance, the well--known $\sigma^2(r)$ measures the
fluctuations of the number of galaxies in a sphere with radius $r$ in
excess of a Poisson process.  $\sigma^2(r)$ is a pure second--order
measure.  To calculate $\sigma^2(r)$ for the PSCz samples, we estimate
the correlation integral $C(r)$, the mean number of points in a sphere
with radius $r$ centred on one galaxy, and use the relation
\begin{equation}
\sigma^2(r) = \frac{C(r)}{\rho\ 4\pi/3\ r^3} -1.
\end{equation}
Our results do not depend on the estimator for $C(r)$
{\citep{kerscher:twopoint}}.
Fig.~\ref{fig:sigma-v100} illustrates that on the two--point level
there are no significant fluctuations between north and south.

Hence, the fluctuations observed with the MFs are a genuine effect of
higher--order correlations, illustrating that low--order statistics
such as the number density or the two--point correlation function may
miss global features of the large--scale structure.  Such
higher--order correlations may originate from coherent elements like
filaments or walls with an extent comparable to or larger than the
size of the sample, in our case 100\hMpc.  This conjecture is also
supported by visual inspection of the large--scale structure (see
e.g.\ {\citealt{huchra:cfa2s1}}).

Other methods to characterise the spatial distribution of points are
\begin{itemize}
\item
the spherical contact distribution $F(r)$, i.e.\ the distribution
function of the distances $r$ of galaxies from an arbitrary point in
space,
\item
the nearest neighbour distance distribution $G(r)$ defined as the
distribution function of distance $r$ of a galaxy to the nearest other
galaxy,
\item
and the ratio suggested by {\citet{vanlieshout:j}}
\begin{equation}
\label{eq:def-J}
J(r) = \frac{1-G(r)}{1-F(r)}.
\end{equation}
\end{itemize}
For a   Poisson distribution one finds
\begin{equation} \label{eq:FG_poi}
G(r) = F(r) = 1 - \exp\left(- \rho \frac{4 \pi}{3} r^3\right)
\end{equation}
and consequently $J(r)=1$.  We use the minus--estimators, as discussed 
by {\citet{kerscher:fluctuations}}, to determine $F$ and $G$ from the 
galaxy distribution.  The spherical contact distribution equals the 
volume density of the first Minkowski functional and one minus the 
void probability: $F(r)=M_0(\CA_r)/|\Omega|=1-P_0(B_r)$.

\begin{figure*}
\begin{center}
\includegraphics[width=5.9cm]{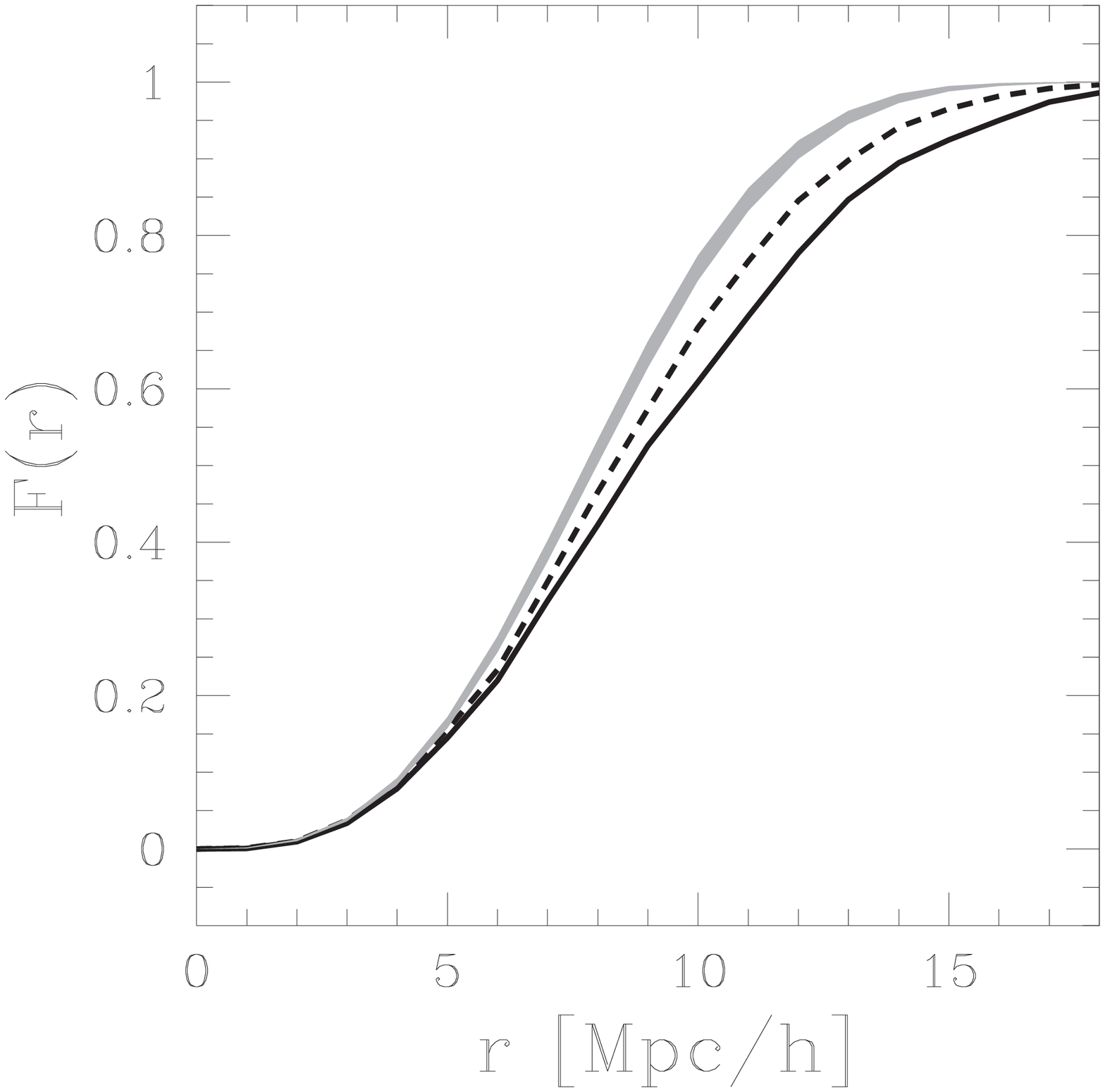}
\includegraphics[width=5.9cm]{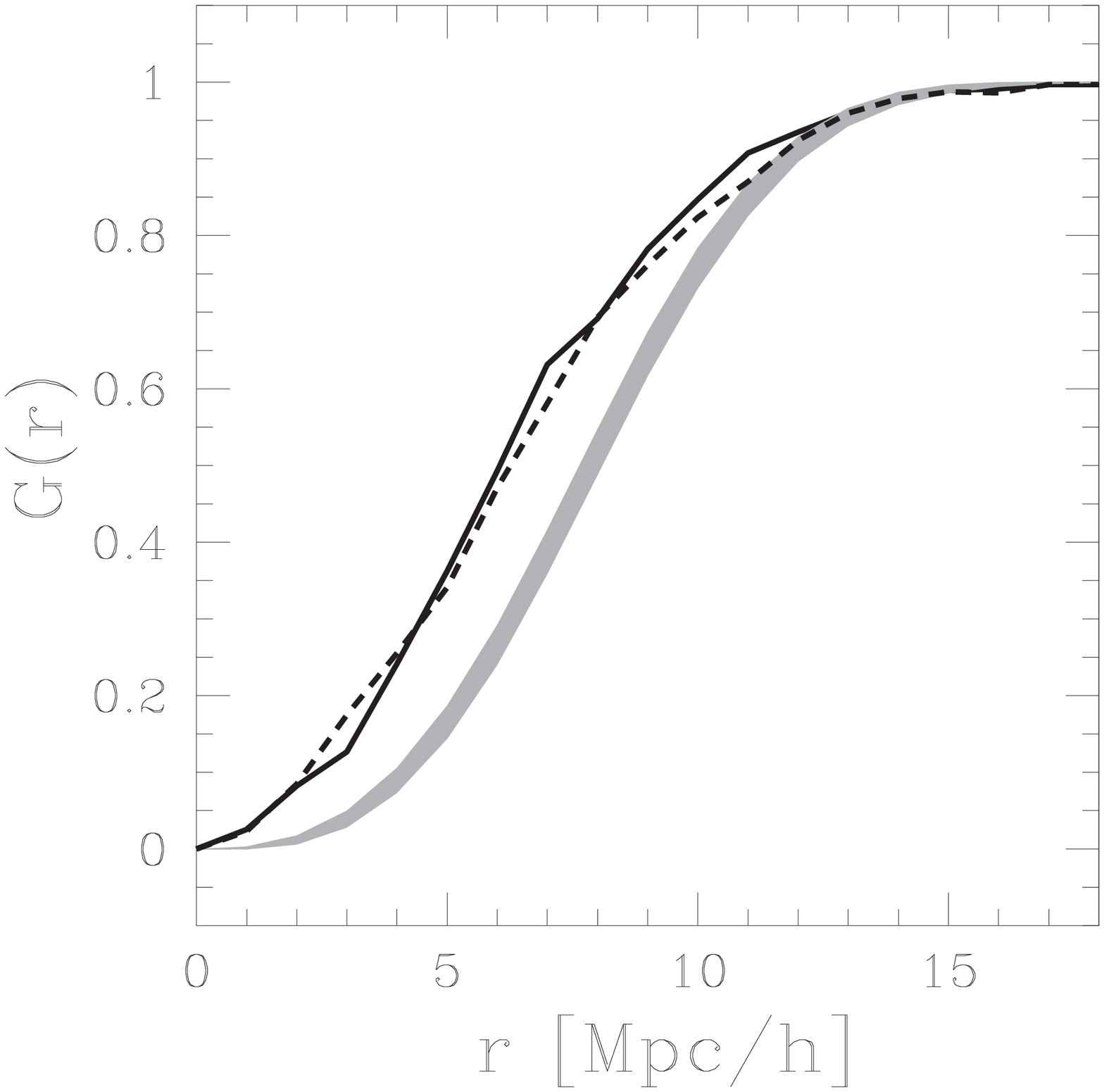}
\includegraphics[width=5.9cm]{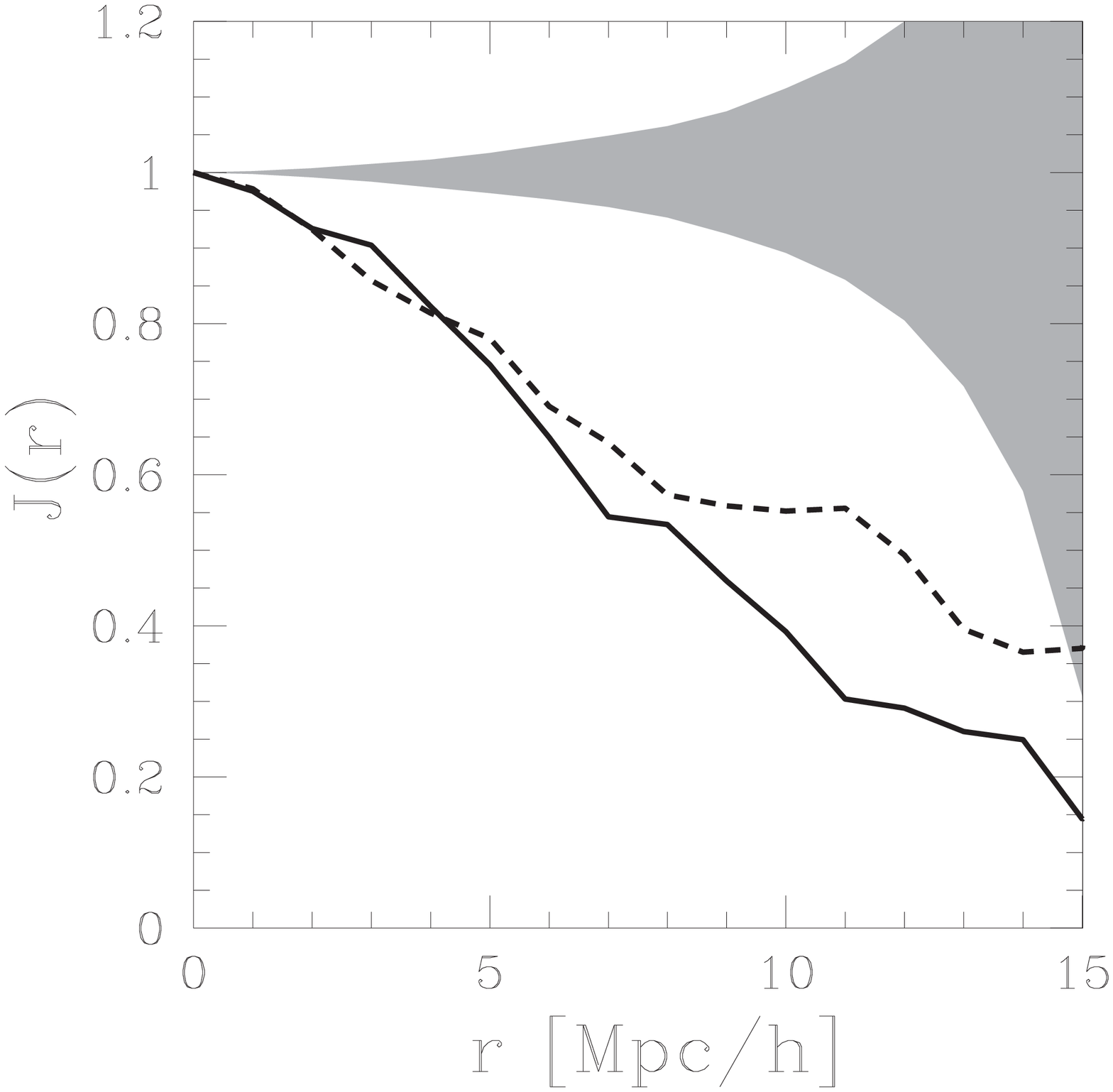}
\end{center}
\caption{
\label{fig:fgj-v100} 
The   panels show  the  $F(r)$,  $G(r)$,  and $J(r)$  function  of
volume--limited  samples from  the PSCz  survey with  100\hMpc\ depth.
Plotting  styles are  the  same as  in Fig.~\ref{fig:MF-v100}.}
\end{figure*}

In Fig.~\ref{fig:fgj-v100} we show the results for the spherical
contact distribution $F(r)$, the nearest neighbour distribution
$G(r)$, and the $J(r)$ function.  The fluctuations within the PSCz
survey in comparison to a Poisson process are most prominent in the
spherical contact distribution $F(r)$, and slightly less pronounced in
the $J(r)$--function.  Virtually no fluctuations are detected with the
nearest neighbour distance distribution $G(r)$.  The reason is that
$G(r)$ focuses on the small--scale features, which do not fluctuate
strongly on the scales probed by the PSCz survey, while $F(r)$ and
$J(r)$ are capable of tracing the large--scale geometry and topology
of the galaxy distribution.

\subsection{From small to large scales}
\label{sec:small-large}

In the preceding sections we saw that the morphology on small scales
up to $10\hMpc$ does not show the strong fluctuations we observed on
larger scales.  To understand this behaviour in more detail we expand
the Minkowski functionals $\Phi_\mu(\CA_r)$ as a series in
$\eta=M_0(B_r)\rho$ around zero.  Based on the expansion in terms of
$n$--point densities ({\citealt{mecke:robust}}, Eq.~(25)) we
obtain
\begin{multline}
\label{eq:eta-expansion}
\phi_\mu(\CA_r) = 1 - \\
\eta\frac{3}{2r^3} \int_0^{2r}\rmd s s^2
\tfrac{M_\mu(B_r(0)\cap B_r(s))}{M_\mu(B_r)} \left[1+\xi_2(s)\right] +
\CO(\eta^2) \;.
\end{multline}
The integral together with the ratio in front yields a dimensionless
geometric factor depending on the correlation function $\xi_2(s)$.
The higher--order terms proportional to $\eta^{n-1}$ include integrals
with intersections of $n$ spheres weighted by the $n$--point
densities.  For small $\eta$ only the two--point correlation function
is important, but the MFs become increasingly more sensitive to
higher--order correlations with larger $\eta$.
A small $\eta$ may be obtained either from a low number density $\rho$
or a small radius $r$ of the spheres.

The two--point correlation function of the PSCz survey shows only tiny
fluctuations between north and south as can be deduced from
Fig.~\ref{fig:sigma-v100}.  On small scales with $\eta\ll1$ the
numerical values of the Minkowski functionals may be reproduced using
Eq.~\eqref{eq:eta-expansion} with the observed two--point
correlation function and mean density.  This explains why for small
radii the morphological fluctuations are negligible.
The expression~\eqref{eq:eta-expansion} also explains why the MFs of
dilute samples tend to look more like those of a Poisson process.

\section{Comparing data and models}
\label{sec:comparison}

\subsection{Second-order moments of Minkowski functionals}

To compare the fluctuations of the observed Minkowski functionals with
different models we will use the dimensionless second--order moments
(see also Eq.~\eqref{eq:vschw2})
\begin{equation}
m_{\mu\mu}(\eta)  
= \frac{M_0(B_r)}{M_\mu(B_r)^2 |\Omega|} 
\left\langle \Big(M_\mu(\CA_r)- \langle M_\mu(\CA_r)\rangle \Big)^2 \right\rangle,
\end{equation}
where $|\Omega|$ is the sample volume.  
The $m_{\mu\mu}$ are estimated either on the base of samples from
simulations or from the real data using the standard estimator for a
variance.
\begin{equation}
m_{\mu\mu}(\eta)  = \frac{M_0(B_r)}{M_\mu(B_r)^2 |\Omega|}\ 
\frac{1}{N_s-1}\sum_{i=1}^{N_s}\Big(M_\mu^{(i)} - M_\mu\Big)^2 ,
\end{equation}
where $N_s$ is the number of samples,
$M_\mu=\frac{1}{N_s}\sum_{i=1}^{N_s}M_\mu^{(i)}$ is the sample mean,
and the $M_\mu^{(i)}$ are the MFs of the $i$th sample. From the PSCz
catalogue we take two samples, the southern and the northern part with
a depth of 100\hMpc\ each.

\begin{figure*}
\begin{center}
\includegraphics[width=5.9cm]{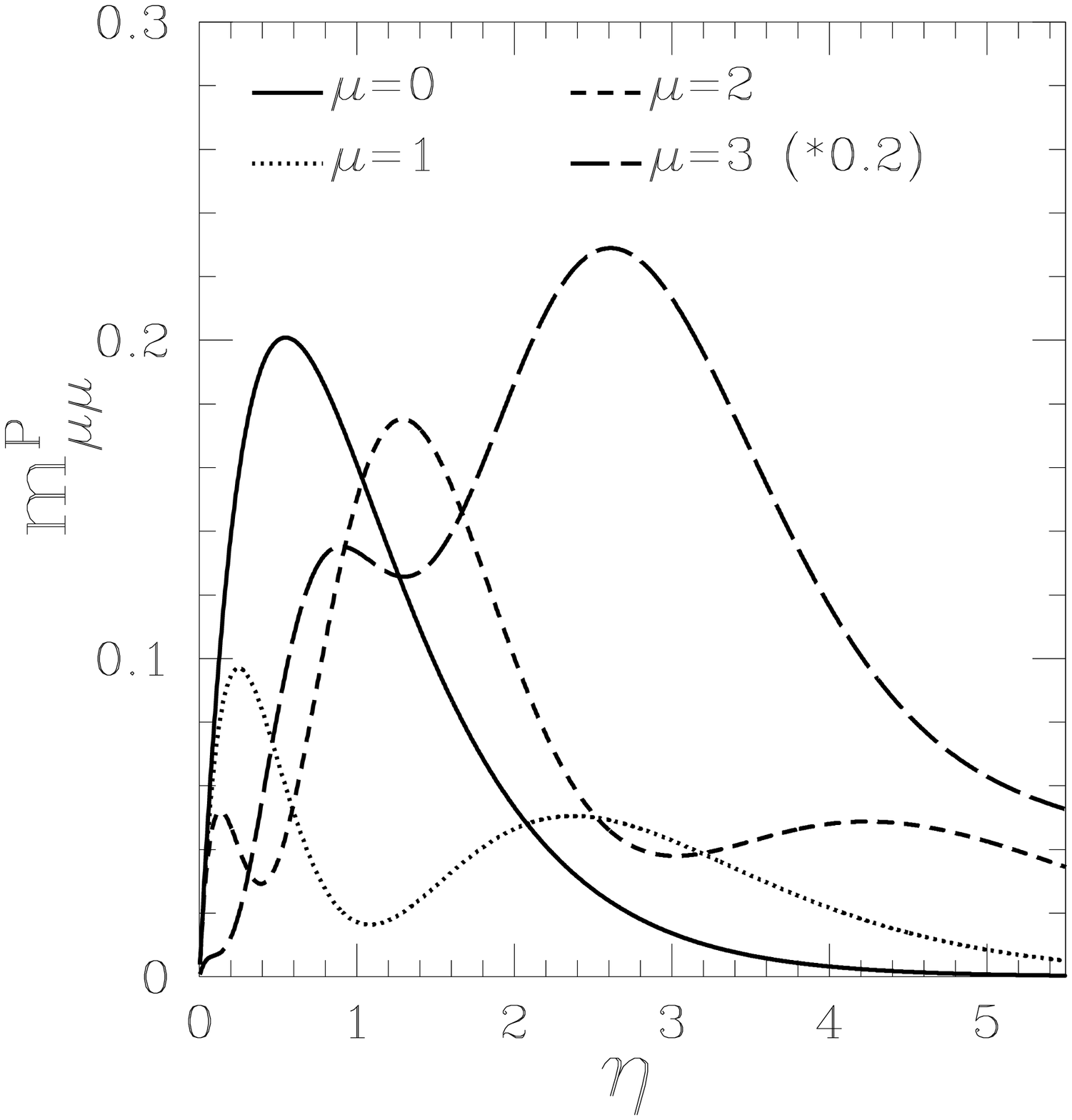}
\includegraphics[width=5.9cm]{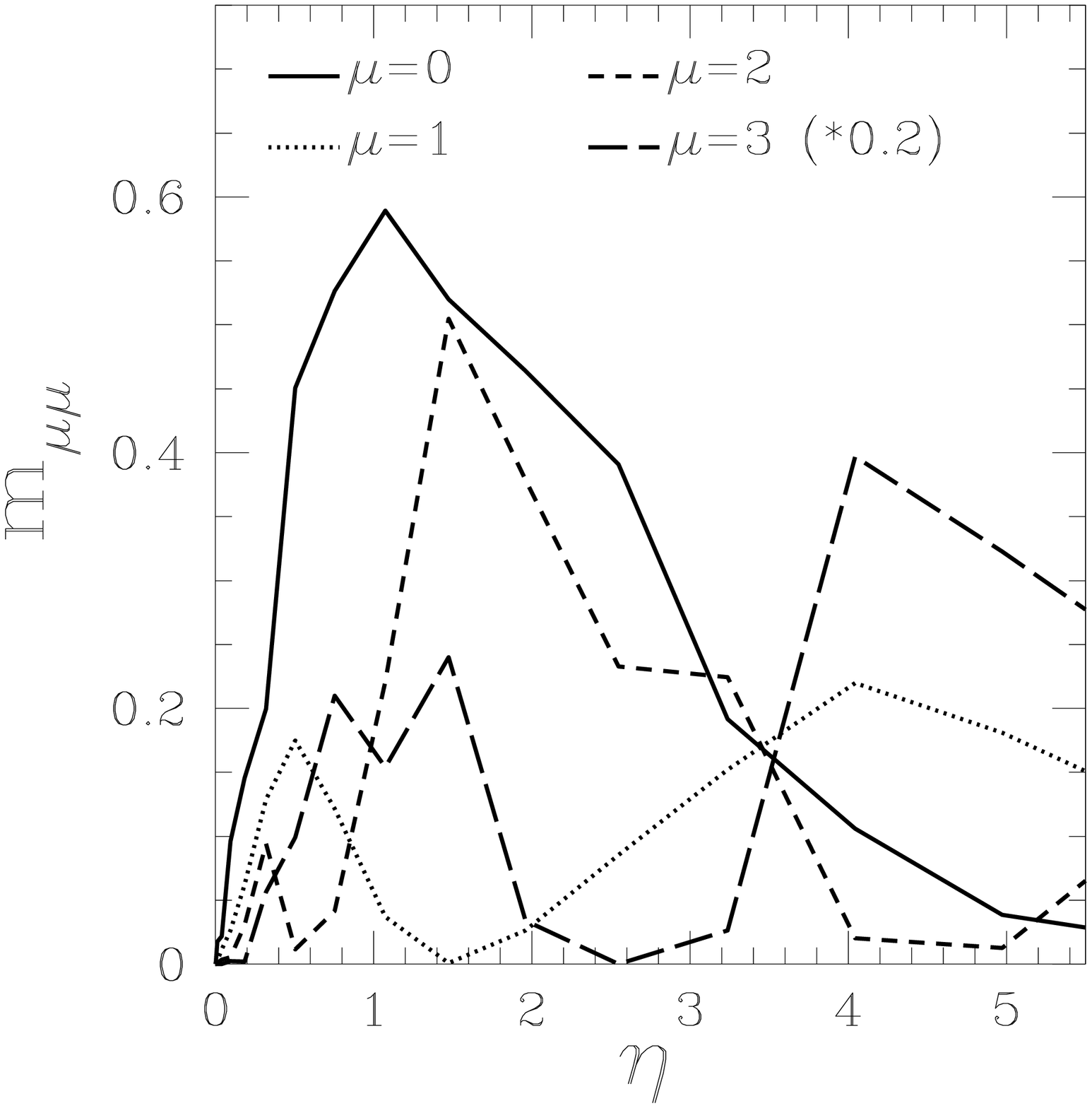}
\includegraphics[width=5.9cm]{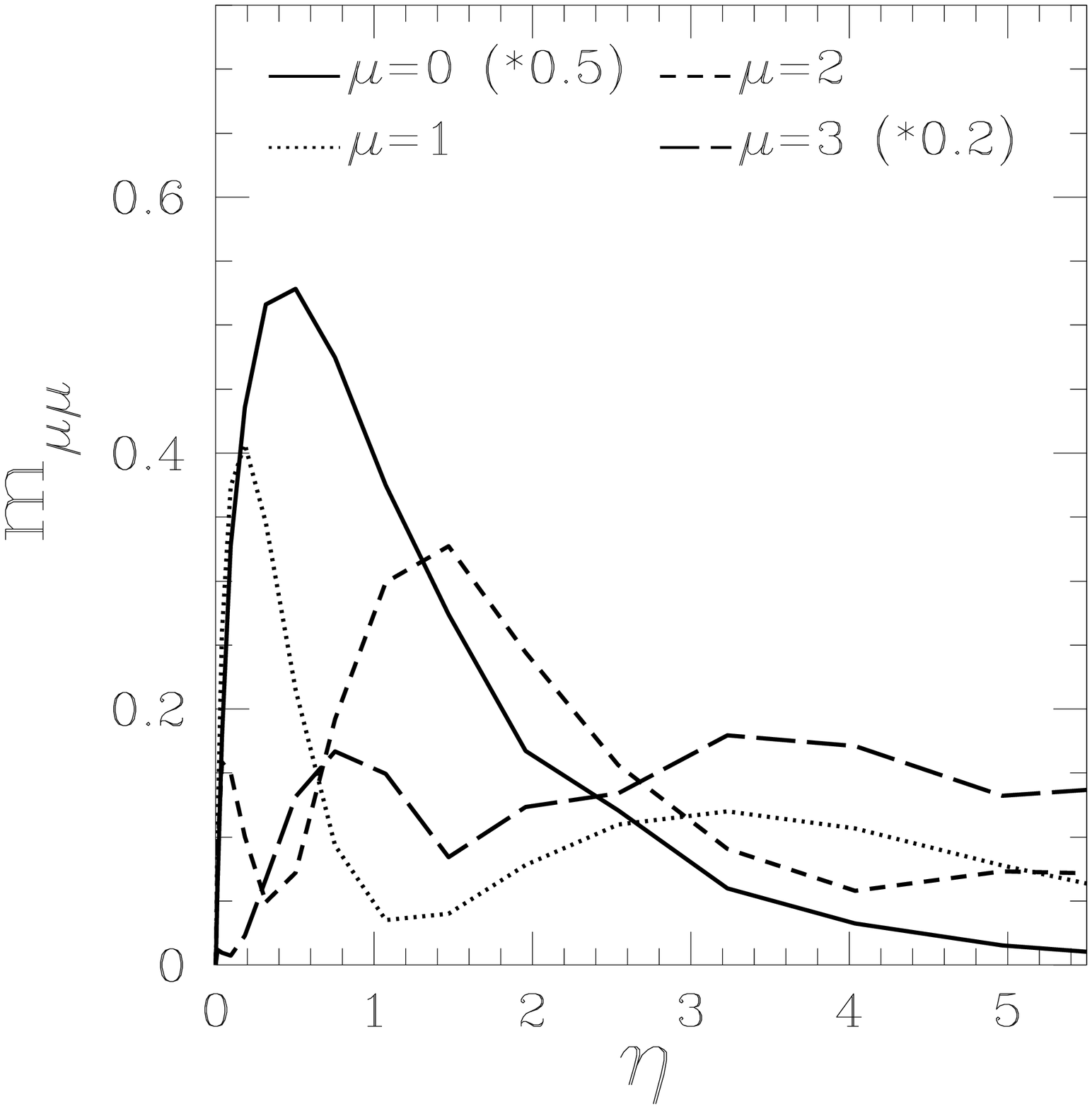}
\end{center}
\caption{
\label{fig:variance}
The dimensionless fluctuations of the Minkowski functionals for a
Poisson process are shown in the left plot.  In the middle plot the
estimated variances $m_{\mu\mu}$ from the volume--limited
sample of the PSCZ with 100\hMpc\ depth, and in the right plot the
estimated variances from the $\Lambda$CDM mock samples are shown
(notice the different scaling).  The $\tau$CDM mock samples show similar
fluctuations.}
\end{figure*}

In Appendix~A the fluctuations for a Poisson and a binomial process
are given analytically.  The comparison with the estimated
fluctuations from the PSCz survey in Fig.~\ref{fig:variance}
indicates that the observed dimensionless fluctuations are
significantly larger than the $m_{\mu\mu}^{\text{P}}$ for a Poisson
process with the same number density.  Qualitatively, the fluctuations
from the PSCz survey show similar features as for a Poisson process,
i.e.\ the functional form and the relative strengths of the
fluctuations are comparable.  Only the surface area is qualitatively
different on large scales.  Of course, an estimate of fluctuations
from two samples only should be considered as a first rough estimate.
Also the zero variance determined from the PSCz for some radii is
an artifact of using only two samples.

\subsection{Comparison with CDM simulations}

More realistic models to compare our data to are based on $N$--body
simulations.  We use mock galaxy catalogues constructed from the VIRGO
consortium's Hubble volume simulations. The mock catalogues were
kindly provided to us by Shaun Cole.  From these mock catalogues we
extract volume--limited samples with a depth of 100\hMpc\ similar to
the observed galaxy samples.
The CDM $N$--body simulations were performed by the VIRGO consortium and
comprise a $\Lambda$CDM ($\Omega_m=0.3$, $\Omega_\Lambda=0.7$,
$\sigma_8=0.9$) and a $\tau$CDM ($\Omega_m=1$, $\Omega_\Lambda=0$,
$\sigma_8=0.6$) simulation, conducted in a box with a side length of
$3h^{-1}$Gpc and $2h^{-1}$Gpc, respectively (for details see
{\citealt{evrard:hubblevolume}} and their web--site\footnote{\tt
http://www.physics.lsa.umich.edu/hubble-volume/}).
Mock catalogues were generated with methods similar to those used by
{\citet{cole:mock}}: the galaxies were extracted from the distribution
of the dark matter particles employing a biasing scheme; luminosities
have been assigned according to the observed luminosity function (for
details see {\citealt{cole:mock}} and their web--site\footnote{\tt
http://star-www.dur.ac.uk/\~{}cole/mocks/hubble.html}).  The observer
position was not chosen at random but selected to satisfy criteria
similar to those of the local group, with respect to the local
density, the CMB dipole, and the local shear {\citep{baugh:mock}}.  We
used the first eleven mock catalogues constructed from each
simulation.

\begin{figure*}
\begin{center}
\includegraphics[width=5.9cm]{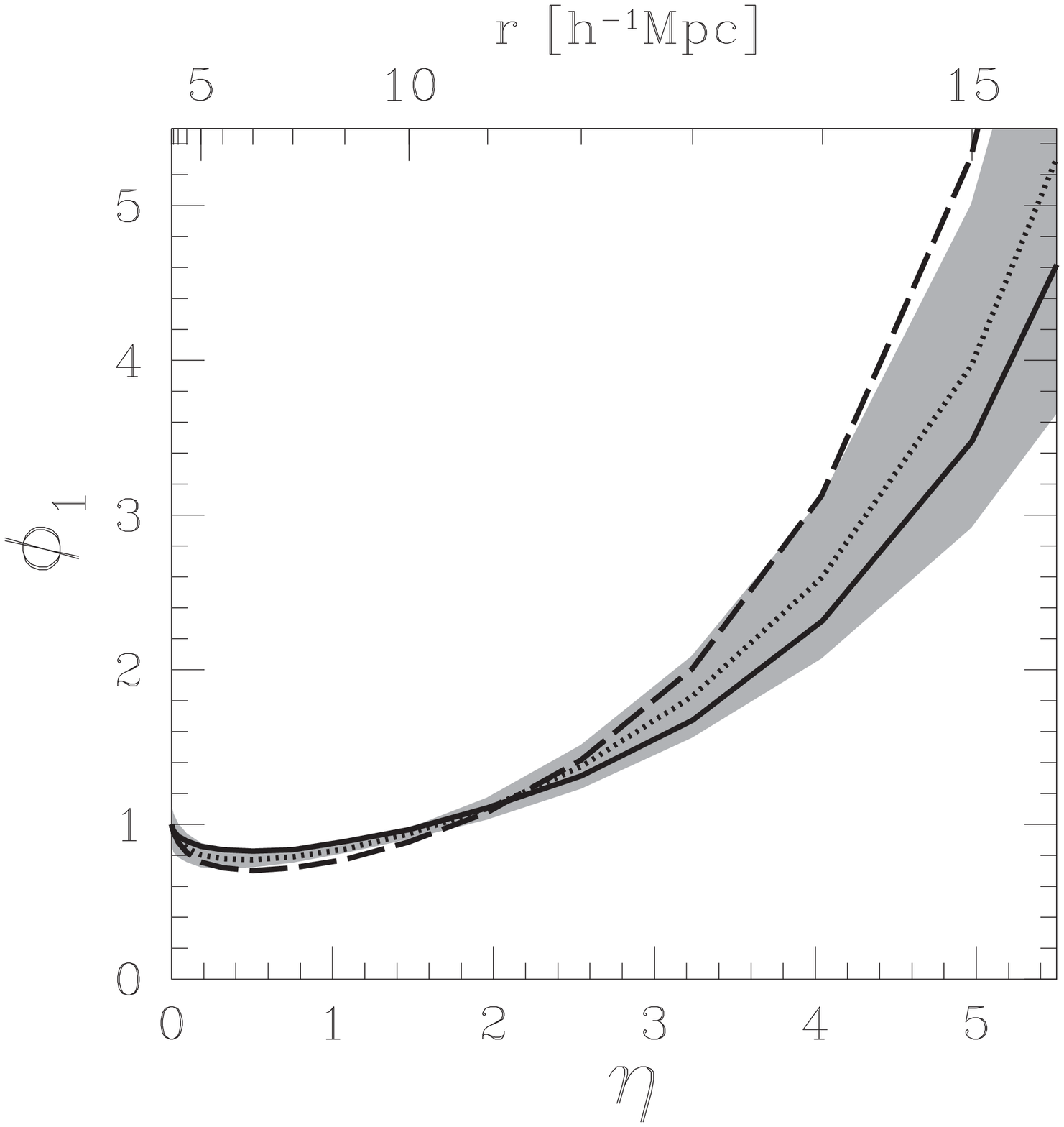}
\includegraphics[width=5.9cm]{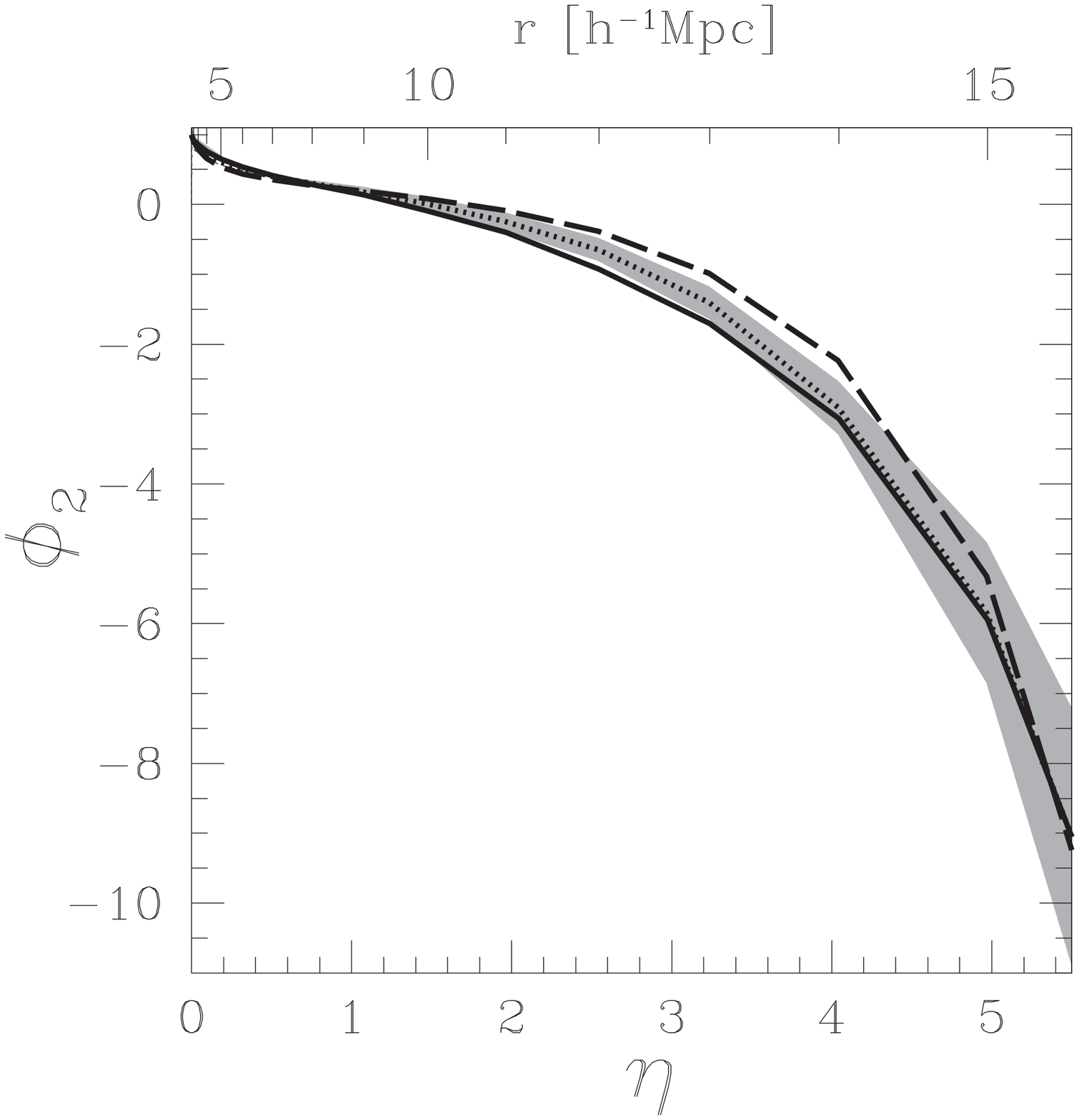}
\includegraphics[width=5.9cm]{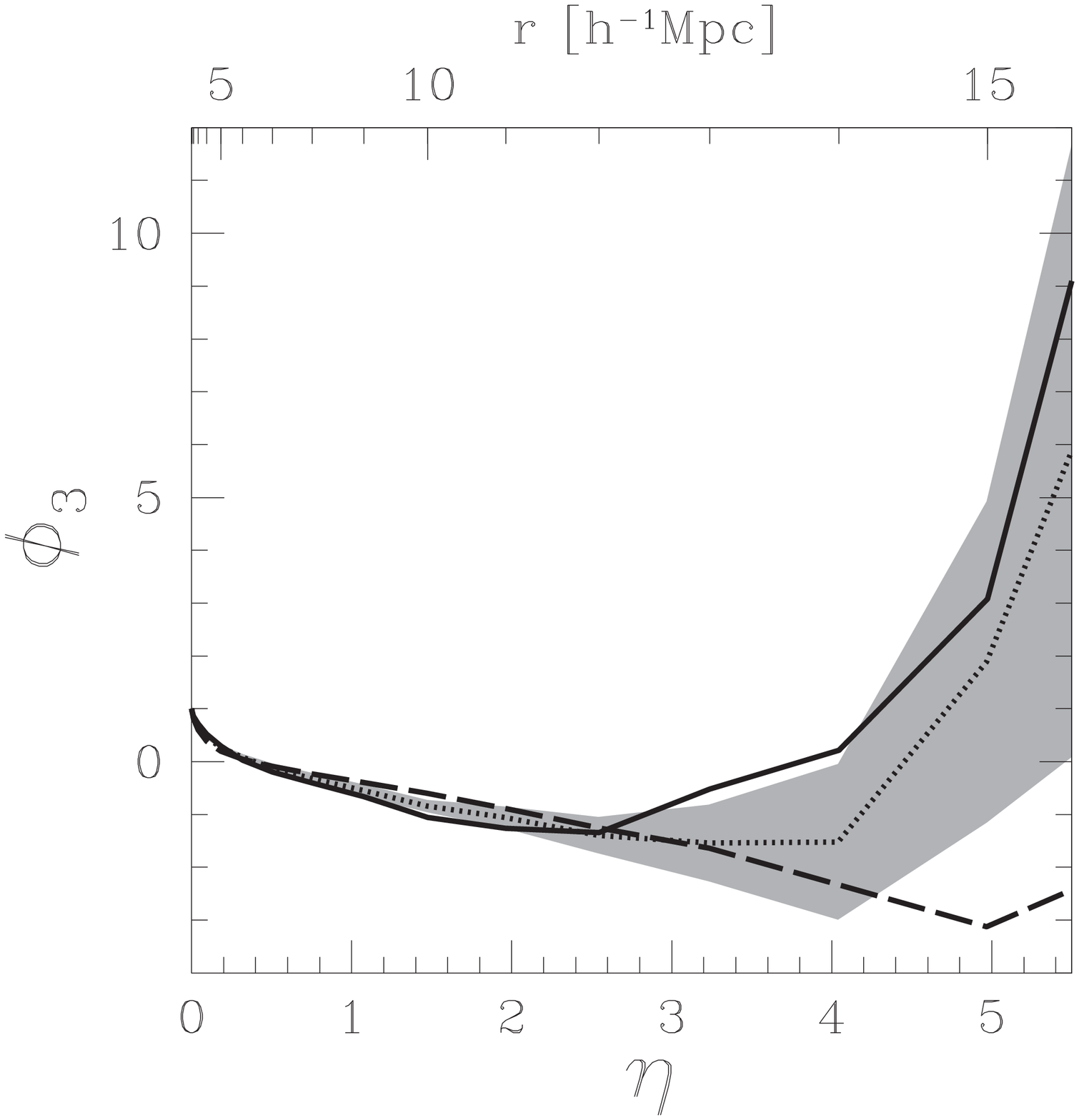}
\end{center}
\caption{
\label{fig:MF-simulation}
In addition to the mean MFs from the PSCz survey (solid), this plot displays
the mean results from the mock sample constructed from the
$\Lambda$CDM (dotted) and $\tau$CDM (dashed) simulation.  The shaded area
marks the one--sigma region determined from the $\Lambda$CDM
simulation.  The $\tau$CDM simulation shows similar fluctuations.}
\end{figure*}

The Minkowski functionals for the mock samples were calculated in
redshift space within the same sample geometry as for the PSCz survey.
Considering north and south separately we can use $N_s=22$ samples per
simulation to estimate the mean MFs and their variance.  In
Fig.~\ref{fig:MF-simulation} the mean MFs of the mock samples are
compared with the mean MFs of the PSCz catalogue.  Within the
estimated errors, the large scale morphology of the galaxy
distribution is marginally reproduced by the $\Lambda$CDM model,
whereas the $\tau$CDM model is too lumpy.

The main issue is whether these mock catalogues are able to reproduce
the large morphological fluctuations in the PSCz survey.  This is of
specific interest since it turned out that a simulation, designed to
match the local distribution of IRAS galaxies, was not able to
reproduce the fluctuations in the 1.2Jy galaxy survey
{\citep{kerscher:fluctuations}}.
Similar large--scale fluctuations as observed in the galaxy
distribution may indeed be found in the mock catalogues of both models
(see Fig.~\ref{fig:variance}).  At least, the amplitude of the
fluctuations $\widehat{m}_{\mu\mu}$ determined from real data and the
mock samples are of the same order of magnitude.  The fluctuations in
the number density between of the volume--limited samples extracted
from the mock catalogues are in full agreement with the $\sigma^2$
expected for the models at the sample size, but are larger than the
number density fluctuations between the northern and southern PSCz
samples.  These fluctuations in the number density seem to cause the
increased fluctuations in $\widehat{m}_{00}$ for the mock samples.

\section{Summary}
\label{sec:summary}

We investigate the morphology of the galaxy distribution in the PSCz
survey.  In order to quantify the geometrical and topological
features, we use the fluctuations of count--in--cells as a two--point
measure, as well as a number of higher--order statistics, namely the
nearest neighbour distance distribution function, the void
probability, the $J$--function, and the family of scalar Minkowski
functionals.  We find appreciable fluctuations between the northern
and southern parts of the survey, confirming earlier findings from the
IRAS 1.2Jy catalogue {\citep{kerscher:fluctuations}}.  These
fluctuations are discernible in volume--limited samples of as much as
200\hMpc\ depth.  By inspecting subsamples we show that these
fluctuations are not due to a specific north--south anisotropy, but
are an intrinsic feature of the galaxy distribution on large scales.

The comparison with the fluctuations in mock--catalogues shows that
$N$--body simulations of two CDM models in boxes with a side length
of $2$ and $3h^{-1}$Gpc followed by a suitable biasing procedure are
able to account for the observed morphological fluctuations in the
galaxy distribution.  This was not the case when we compared the
fluctuations in the IRAS 1.2Jy with those determined from a simulation
in a smaller box with 250\hMpc\ side length.  Concerning the mean
values of the Minkowski functionals, the $\tau$CDM simulation is not
able to explain the observed large--scale morphology.  Marginally, the
$\Lambda$CDM model succeeds.

Speaking of  fluctuations, one usually  thinks of the  fluctuations in
the  number   of  galaxies,  the  fluctuations  of   count--in--cells
$\sigma^2$,  or   the  power   spectrum  quantifying  the   amount  of
fluctuations in a Fourier mode of the density contrast.  Both measures
quantify  the galaxy  distribution at  the two--point  level,  and are
therefore  unable to detect coherent large--scale  structure elements
like        walls        and        filaments        (see        e.g.\
{\citealt{szalay:walls,schmalzing:disentanglingI,chiang:phase}}).
Indeed,  our investigations  of the  galaxy distribution  in  the PSCz
survey  using  $\sigma^2$  show  no significant  fluctuations  at  the
two--point level between the northern and southern parts of the galaxy
sample.

Minkowski functionals, however, incorporate correlations of arbitrary
order and quantify the connectivity of large--scale structure elements
in the galaxy distribution.  The amplitude of the morphological
fluctuations in samples with an extent of 100\hMpc\ is approximately
three times larger than expected for a Poisson process. The
morphological fluctuations remain detectable even in a 200\hMpc\
sample, but since the deeper volume--limited samples are sparser and
therefore more similar to a Poisson process, the results become less
significant.  The structural differences between the northern and
southern parts of the PSCz sample show up prominently in the Minkowski
functionals for radii above 10\hMpc.  As already mentioned, the number
density and also the two--point correlations show only small
fluctuations.  This indicates that a variety of coherent structures
with a low density contrast and an extent of more than 100\hMpc,
perhaps 200\hMpc, are shaping the large--scale structure of the
Universe.  In this sense, our analysis quantifies the visual
impression of features like the great wall in the CfA2 catalogue
{\citep{huchra:cfa2s1}}.

Structures like  walls and filaments were predicted  by analytical and
numerical    work    based    on    the    Zel'dovich    approximation
{\citep{arnold:largescale,doroshkevich:superlargeII}}    and   related
approximations   {\citep{kofman:coherent,bond:filament}}.    They  are
generic features  of the gravitational collapse  for CDM--like initial
conditions,   as   numerous   $N$--body   simulations   could   verify
{\citep{melott:generation,jenkins:evolution}}.     To   detect   these
morphological  features, especially  if they  show  up only  at a  low
density contrast, sensitive methods like the Minkowski functionals are
needed.  They  allow us to quantify  the geometry and  topology of the
network of  walls, filaments,  and voids seen  in the  distribution of
galaxies.

\section*{Acknowledgements}

We would like to thank the VIRGO consortium and the Durham
astrophysics theory group for providing us the Hubble volume
simulations and the mock catalogues. We thank Alvaro Dom\'\i nguez for
helpful comments on the manuscript. Special thanks go to Shaun Cole
for constructing the desired mock samples and for clarifying comments.

CB, TB, and MK acknowledge support from the {\em
Sonderforschungsbereich 375 f{\"u}r Astroteilchenphysik der DFG}. MK
acknowledges support from the NSF grant AST~9802980.  KM acknowledges
support from the DFG grant ME1361/6-1.  TB acknowledges support and
hospitality by the National Astronomical Observatory in Tokyo, as well
as hospitality at Tohoku University in Sendai, Japan, and Geneva
University with support by the Tomalla Foundation, Switzerland.  This
work was supported by Danmarks Grundforskningsfond through its support
for TAC.


\onecolumn
\appendix
\section{Fluctuations of the Minkowski functionals for a Poisson model}
\label{sec:analytical-fluctuations}

The simplest point process, and thus the simplest reference model to
compare real data to, is a stationary Poisson point process lacking any
spatial correlations.  In this appendix we give the exact expressions
for the variances of the Minkowski functionals for a Poisson process
and a binomial process.  These results prove useful both to estimate
the fluctuations of the morphological properties for finite samples,
and to serve as a reference model for a comparison with fluctuations in
empirical datasets (Fig.~\ref{fig:variance}).  Here we only explain
the basic notions and list the results for three dimensions.  A full
derivation of the variances for the Minkowski functionals as well as
their covariances in any dimension can be found in \citep{mecke:exact}.

Let us first consider a stationary Poisson point process with number
density $\rho$ in a sample $\Omega$ of volume $|\Omega|$ embedded into
three--dimensional Euclidean space.  We are interested in the
morphological fluctuations as quantified by the covariances of the
Minkowski functionals $M_\mu(\CA_r)$ for the Boolean grain model.  For
convenience, we divide the variances by the Minkowski functionals
$M_\mu(B_r)$ of a single sphere $B_r$ of radius $r$ and express them
as functions of  $\eta \equiv M_0(B_r)\rho$ such
that all quantities are dimensionless and may be compared
quantitatively:
\begin{align}
m_{\mu_1\mu_2}(\eta)  
\equiv
\frac{M_0(B_r)}{|\Omega|M_{\mu_1}(B_r)M_{\mu_2}(B_r)} 
\Bigg\langle
\Big(M_{\mu_1}(\CA_r)-\big\langle M_{\mu_1}(\CA_r)\big\rangle\Big)
\Big(M_{\mu_2}(\CA_r)-\big\langle M_{\mu_2}(\CA_r)\big\rangle\Big)
\Bigg\rangle,
\end{align}
where $\langle\cdot\rangle$ denotes the average over the statistical
ensemble.
  
The Minkowski functionals $M_\mu(\CA_r)$ of a union set of balls can
be represented either as an integral over the volume of the set
($\mu=0$) or as integrals over the $(d-\lambda)$--dimensional
intersections of $\lambda$ spherical boundaries for $\mu=1,\ldots,d$
and $\lambda=1,\ldots,\mu$; in the latter case local curvatures arise
as weightings.  Therefore, the variances $m_{\mu_1\mu_2}$ can be
decomposed into a series of {\em curvature--weighted structure
functions} $M_{\mu_1\mu_2}^{\lambda_1\lambda_2} (\eta,s)$ containing
the contributions of the $(d-\lambda_i)$--dimensional spherical
intersections to $M_{\mu_i}$ for $i=1,2$ \citep{mecke:exact}:
\begin{equation}\label{eq:vschw2}
m_{\mu_1\mu_2}(\eta)  
=
\sum_{\lambda_1=0}^{\mu_1} \sum_{\lambda_2=0}^{\mu_2}
\int\limits_0^1\rmd s\ M_{\mu_1\mu_2}^{\lambda_1\lambda_2}(\eta;s).
\end{equation}    
For the following listing of the structure functions $V$ denotes the
volume of the sphere $B_r$, and $V(s)$ is the normalized volume of two
overlapping spheres $B_r(\bx_1)$ and $B_r(\bx_2)$ of radius $r$ and
centre $\bx_i$ given by
\begin{equation}
V(s)
=
\frac{V(B_r(\bx_1)\cup B_r(\bx_2))}{V(B_r)}
=    
1+\frac{3}{2}s-\frac{1}{2}s^3,   
\end{equation}
where $s=\frac{\left|\bx_2-\bx_1\right|}{2r}$ denotes the normalized
distance of the spheres.  Note, that the structure functions are
symmetric, $M^{\lambda_1\lambda_2}_{\mu_1\mu_2}(\eta,s) =
M^{\lambda_2\lambda_1}_{\mu_2\mu_1}(\eta,s)$, and vanish, if
$\lambda_i>\mu_i$ for $i=1$ or $i=2$. We get
\begin{gather}
\label{eq:difference2s} 
M_{00}^{00}(\eta;s)  
 = 24 s^2 \left(\rme^{-\eta V(s)}  - \rme^{-2\eta }\right) , \\  
M_{\mu 0}^{\lambda 0}(\eta;s)  
 =  (-1)^{\lambda+1}\frac{\eta^\lambda}{\lambda!}  
\left(\bar{M}^{\lambda}_\mu  \rme^{-2\eta } - \check{M}^{\lambda}_\mu 
\rme^{-\eta V(s)} \right) ,  \nonumber \\ 
M_{\mu_1\mu_2}^{\lambda_1\lambda_2}(\eta;s)  
 = (-1)^{\lambda_1+\lambda_2} \rme^{-\eta  V(s)} 
\sum\limits_{l=0}^{\lambda_2}
\frac{\eta^{\lambda_1+\lambda_2-l}}{l!(\lambda_1-l)!(\lambda_2-l)!}
M^{\lambda_1\lambda_2}_{\mu_1\mu_2;l}  -
24s^2\; M_{\mu_1}^{\lambda_1}M_{\mu_2}^{\lambda_2}  . \nonumber
\end{gather}
The {\em structure amplitudes} $\bar{M}_\mu^\lambda$, $M_\mu^\lambda (\eta)$,
$\check{M}^{\lambda}_\mu (s)$  and $M^{\lambda_1\lambda_2}_{\mu_1\mu_2;l} (s)$
arising    in   these   formulae    vanish   if    $\lambda>\mu$   and
$\lambda_i>\mu_i$  for $i=1$  or $i=2$,  respectively.  For  the other
cases one  finds the following:  the quantities $\bar{M}^\lambda_\mu$,
with
\begin{equation}
\begin{split}
\bar{M}^1_\mu &= 1 ,\\
\bar{M}^2_2   &= \frac{2M_1(B_r)^2}{M_0(B_r)M_2(B_r)} = \frac{3\pi^2}{16} ,\\
\bar{M}^2_3   &= \frac{6M_1(B_r)M_2(B_r)}{M_0(B_r)M_3(B_r)} = 6 ,\\
\bar{M}^3_3   &= \frac{6M_1(B_r)^3}{M_0(B_r)^2M_3(B_r)} = \frac{9\pi^2}{16}\end{split}
\end{equation}
contain local curvature contributions.  The $M_\mu^\lambda(\eta)$
additionally incorporate the probability of finding $\lambda$ points
within an otherwise empty sphere $B_r$ resulting in an intersection
which -- according to the additivity of the Minkowski functionals --
has to be weighted with $(-1)^{\lambda+1}$: $M_\mu^\lambda(\eta) =
(-1)^{\lambda+1}\frac{\eta^\lambda}{\lambda!}  \rme^{-\eta }
\bar{M}^{\lambda}_\mu$.  The $\check{M}^{\lambda}_\mu (s)$ are given by
\begin{equation}
\begin{split}
\check{M}^1_\mu(s)
&=
\frac{1}{2} (1+s),
\\
\check{M}^2_\mu(s)
&=
\frac{1}{2} \intphi\iintx
 \sqrt{1-(x_1x_2+\sqrt{1-x_1^2}\sqrt{1-x_2^2}\cos\phi)^2}\ F_\mu[t] ,
\\
\check{M}^3_3(s) 
&=
\frac{1}{2} \intx \iintphi\iintx    
\Delta[x_1,x_2,x,\phi_1,\phi_2]
\\
&\qquad
\left|\ 
x\sqrt{1-x_1^2}\sqrt{1-x_2^2}\sin(\phi_2-\phi_1) 
-x_1\sqrt{1-x^2}\sqrt{1-x_2^2}\sin\phi_2
+x_2\sqrt{1-x^2}\sqrt{1-x_1^2}\sin\phi_1
\ \right|  .
\end{split}
\end{equation}

Finally, we turn to the $M^{\lambda_1\lambda_2}_{\mu_1\mu_2;l} (s)$.  For
$l=0$, one finds that they split into a product of two functions
already known from above:
\begin{equation}
M_{\mu_1\mu_2;0}^{\lambda_1\lambda_2}(s)
=
24s^2
\check{M}_{\mu_1}^{\lambda_1}(s)\check{M}_{\mu_2}^{\lambda_2}(s).
\end{equation}
For $l\ne0$, the expressions become  more involved:
\begin{equation}
\begin{split}
\label{eq:structure-amplitudes}
M_{\mu_1\mu_2;1}^{11}(s)
&=
2s,
\\
M_{\mu_1\mu_2;1}^{21}(s)
&=
2s\intphi\intx
F_{\mu_1}[t(-s,x,\phi)]\sqrt{1-(-sx+\sqrt{1-s^2}\sqrt{1-x^2}\cos\phi)^2},
\\
M_{\mu_1\mu_2;1}^{22}(s)
&=
\frac{1}{2s} M_{\mu_1 1;1}^{21}(s) M_{\mu_2 1;1}^{21}(s) ,
\\
M_{\mu_1\mu_2;2}^{22}(s) 
&=
\intphi\left|1-(s^2+(1-s^2)\cos\phi)^2\right|
F_{\mu_1}[t(-s,-s,\phi)]F_{\mu_2}[t(-s,-s,\phi)],
\\ 
M_{3\mu_2;1}^{31}(s) 
&=
2s\iintphi\iintx\ \Delta[-s,-s,x,\phi_1,\phi_2] \\&\qquad
\left|
\sqrt{1-s^2}x_2\sqrt{1-x_1^2}\sin\phi_1
-\sqrt{1-s^2}x_1\sqrt{1-x_2^2}\sin\phi_2
-s\sqrt{1-x_1^2}\sqrt{1-x_2^2}\sin(\phi_2-\phi_1)
\right|,
\\  
M_{3\mu_2;1}^{32}(s) 
&=
\frac{1}{2s} M_{31;1}^{31}(s) M_{\mu_2 1;1}^{21}(s),
\\ 
M_{3\mu_2;2}^{32}(s) 
&=
\iintphi\intx \sqrt{1-(s^2+(1-s^2)\cos\phi_1)^2}
\Delta[-s,-s,x,\phi_1,\phi_2]F_{\mu_2}[t(-s,-s,\phi_2-\phi_1)]
\\
&\qquad
\times\left|
(1-s^2)x\sin\phi_1
+s\sqrt{1-s^2}\sqrt{1-x^2}\left(\sin(\phi_1-\phi_2)+\sin\phi_2\right)
\right|,
\\
M_{33;1}^{33}(s) 
&=
\frac{1}{2s} \left(M_{31;1}^{31}(s)\right)^2 ,
\\ 
M_{33;2}^{33}(s) 
&=
\intphi\iintphi\iintx
\Delta[x_1,-s,-s,\phi_1,\phi] \Delta[x_2,-s,-s,\phi_2,\phi]
\\
&\qquad
\times\prod\limits_{i=1}^2\left|
(1-s^2)x_i\sin\phi
+\sqrt{1-s^2}\sqrt{1-x_i^2}\left(\sin(\phi-\phi_i)+\sin\phi_i\right)
\right|,
\\ 
M_{33;3}^{33}(s) 
&=
\frac{1}{2}s(1-s^2)^2\iintphi
\left|\sin(\phi_1-\phi_2)-\sin\phi_1+\sin\phi_2\right|^2
\left(\Delta[-s,-s,-s,\phi_1,\phi_2]\right)^2  
+2\check{M}_\chi^2\delta(\bx_1-\bx_2),
\end{split} 
\end{equation}
where we have used the functions $F_i$:
\begin{equation}
\begin{split}
F_2[t]  = \arcsin t(x_1,x_2,\phi),
\qquad
F_3[t]  = \frac{2 t(x_1,x_2,\phi)}{\sqrt{1-t(x_1,x_2,\phi)^2}},
\\ 
t(x_1,x_2,\phi)  = 
\frac{1}{\sqrt{2}}\sqrt{1-x_1x_2-\sqrt{1-x_1^2}\sqrt{1-x_2^2}\cos\phi}.
\end{split}
\end{equation}
Furthermore, Eq.~(\ref{eq:structure-amplitudes}) employs the
spherical excess $\Delta[x_1,x_2,x,\phi_1,\phi_2]$ given by
l'Huilier's formula:
\begin{equation}
\tan^2\frac{\Delta}{4}=
\tan\frac{ \alpha_1+\alpha_2+\alpha_3}{4} 
\tan\frac{ \alpha_1+\alpha_2-\alpha_3}{4}
\tan\frac{ \alpha_1-\alpha_2+\alpha_3}{4}
\tan\frac{-\alpha_1+\alpha_2+\alpha_3}{4}, 
\end{equation}
where the edges of a spherical triangle are given by
$\alpha_i=2\arcsin t_i$ and
\begin{equation}
t_1  = t(x_2,x,\phi_2),\qquad
t_2  = t(x,x_1,-\phi_1),\qquad
t_3  = t(x_1,x_2,\phi_1-\phi_2).
\end{equation}
And lastly, the pre--factor of the $\delta$--distribution at the end
of Eq.~(\ref{eq:structure-amplitudes}) is given by
\begin{equation}
\begin{split}
\check{M}_\chi^2
&=
\frac{1}{2} \intx\iintphi\iintx
\Delta[x_1,x_2,x,\phi_1,\phi_2]^2
\\
&\qquad
\left|
x\sqrt{1-x_1^2}\sqrt{1-x_2^2}\sin(\phi_2-\phi_1)
-x_1\sqrt{1-x^2}\sqrt{1-x_2^2}\sin\phi_2  
+x_2\sqrt{1-x^2}\sqrt{1-x_1^2}\sin\phi_1
\right|
.
\end{split}
\end{equation}
While the expressions appear somewhat complicated at first glance,
they can be easily evaluated by numerical integration.

\begin{figure}  
\begin{center}
\includegraphics[width=7cm]{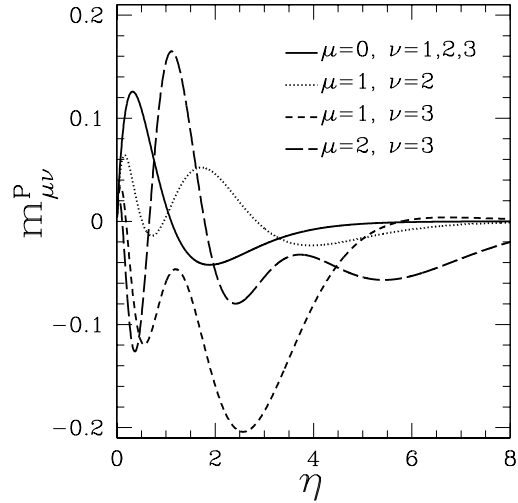}
\end{center}
\caption{
\label{fig:schwankungen3d}   
The normalized  covariances of the Minkowski functionals  of a Poisson
process (see Fig.~\ref{fig:variance} for the variances). }
\end{figure}

In Figs.~\ref{fig:variance} and {}\ref{fig:schwankungen3d} the
second--order moments are shown as functions of 
$\eta=M_0(B_r)\rho$.  One finds an increasing behaviour,
$m_{\nu\mu}(\eta)=\eta+{\cal O}(\eta^2)$, for low densities $\eta$,
and the asymptotic behaviour
$m_{\nu\mu}(\eta\rightarrow\infty)\sim\eta^{\nu+\mu-d}\rme^{-\eta}$.
The Euler characteristic shows by far the largest variances.

\begin{figure*}
\begin{center}
\includegraphics[width=5.9cm]{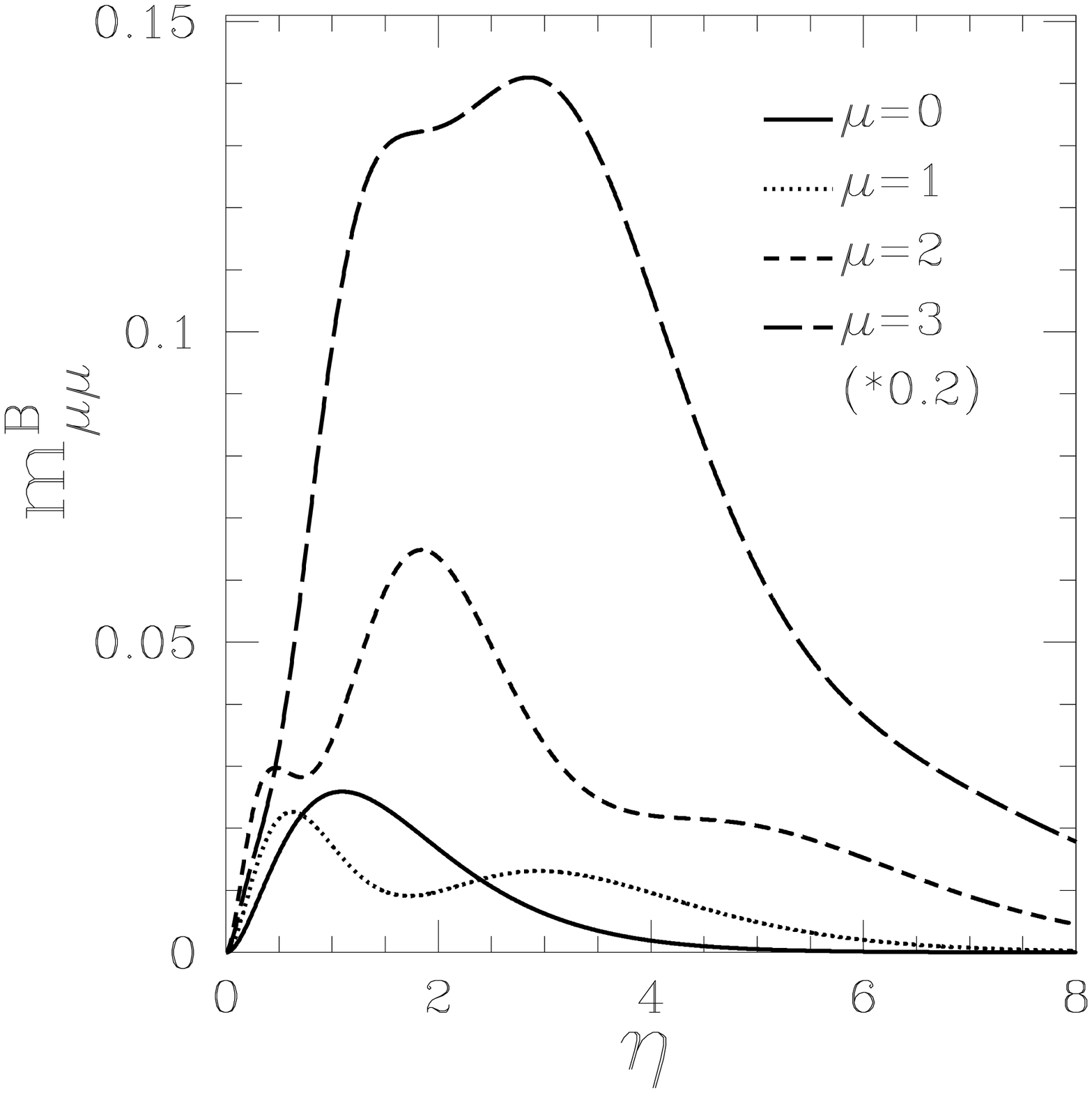}
\includegraphics[width=5.9cm]{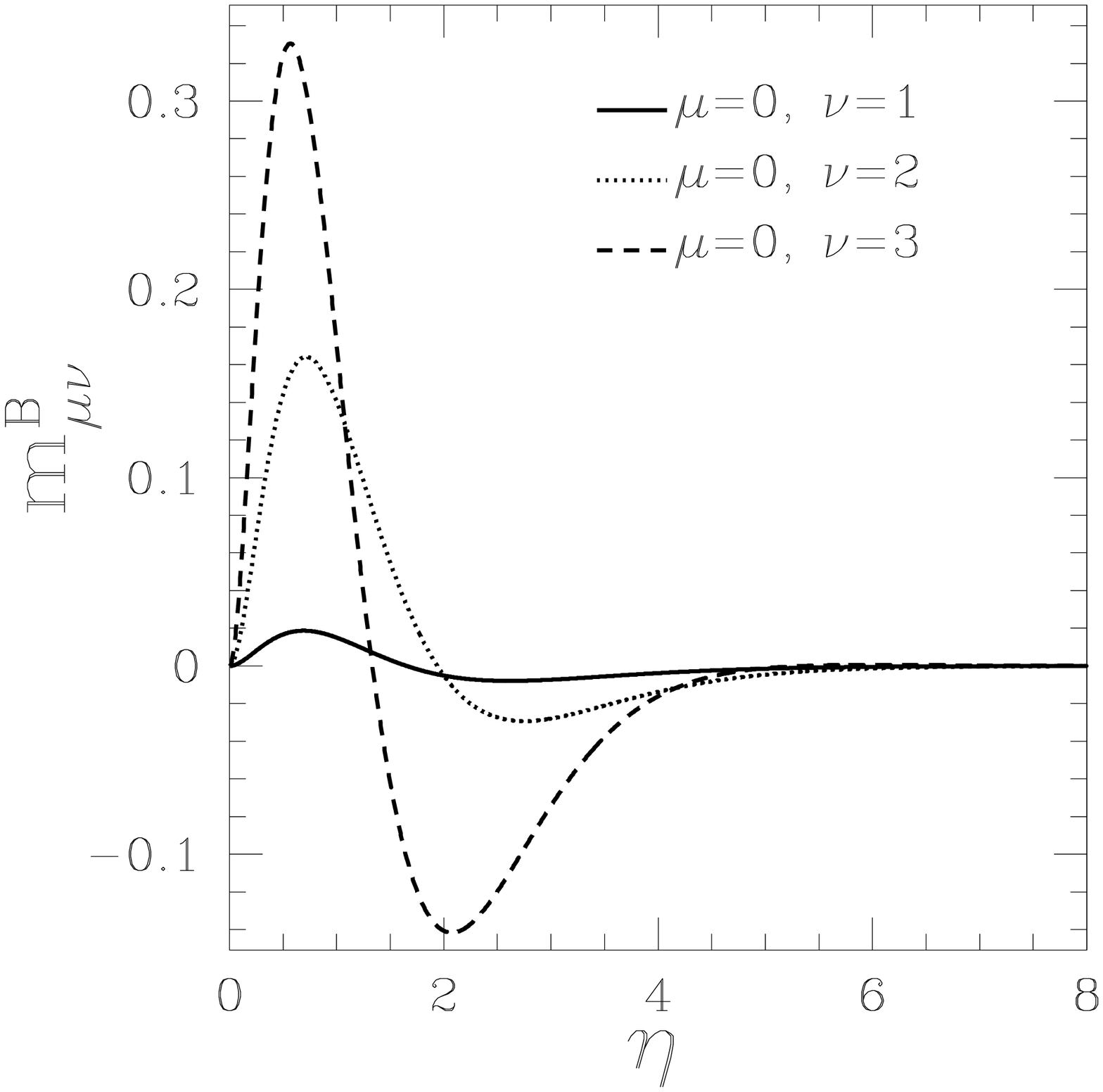}
\includegraphics[width=5.9cm]{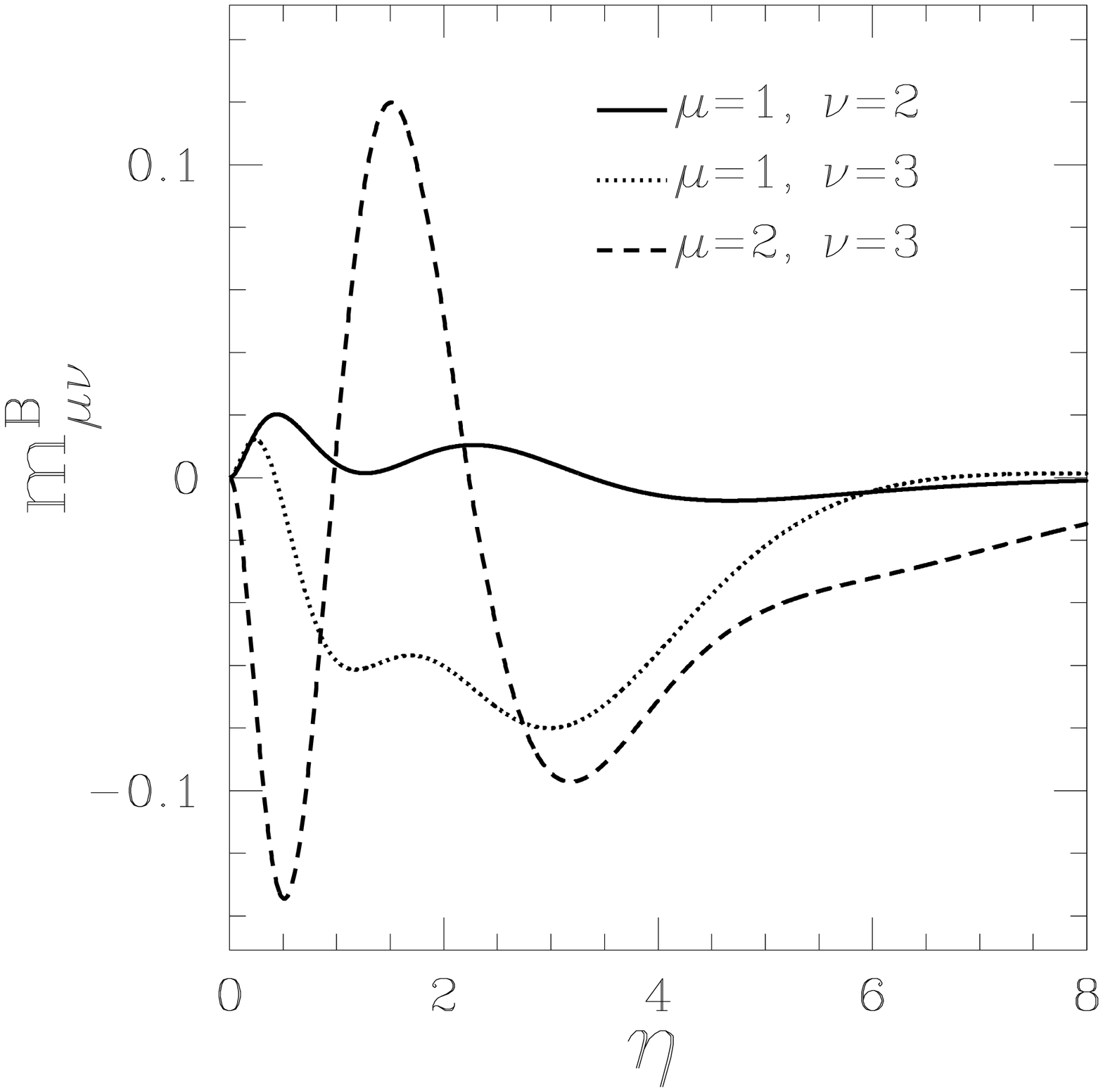}
\end{center}
\caption{
\label{fig:schwankungen3d_2}
The covariances of the Minkowski functionals, now for a binomial
process. }
\end{figure*}

Whereas for a Poisson process (P) the number of points within our
sample is fluctuating, a binomial process (B) corresponds to a
canonical ensemble, where the number of points is fixed.  The
fluctuations of the Minkowski functionals for a binomial process can
be derived from the Poisson case, since one finds that in general
\citep{mecke:exact}
\begin{equation}
m^{\text{P}}_{\mu_1\mu_2}(\eta)-m^{\text{B}}_{\mu_1\mu_2}(\eta) 
=\eta
\frac{\partial\left(\eta\Phi_{\mu_1}^{\text{P}}(\eta)\right)}{\partial\eta} 
\frac{\partial\left(\eta\Phi_{\mu_2}^{\text{P}}(\eta)\right)}{\partial\eta},  
\end{equation}
where $\Phi_{\mu_1}^{\text{P}}(\eta)$ are the mean values given by
Eq.~\eqref{eq:Poisson}. Results for a binomial process are shown
in Fig.~\ref{fig:schwankungen3d_2}.  As shown in
Fig.~\ref{fig:schwankungen3d_2}, the binomial process exhibits
smaller fluctuations compared to the Poisson case depicted in
Fig.~\ref{fig:schwankungen3d}.  The qualitative features in the
fluctuations are similar but significantly damped.  Also additional
correlations between the volume, the surface area, and curvatures show
up.

\end{document}